\documentclass[12pt]{article}
\usepackage[a4paper, margin=1.0in]{geometry}
\usepackage[pdftex]{graphicx}
\usepackage{mathrsfs}
\usepackage{scalerel}
\usepackage{color}
\usepackage{authblk}
\usepackage{psfrag}
\usepackage{graphicx}
\usepackage{caption,subcaption}
\usepackage{enumerate}
\usepackage{amsmath,amssymb,calc,amsfonts}
\usepackage{latexsym}
\usepackage{textpos}
\def\beq{\begin{eqnarray}}
\def\eeq{\end{eqnarray}}

\def\={\stackrel{\Delta}{=}}

\DeclareMathOperator\erf{Erf}
\usepackage[utf8]{inputenc}
\usepackage{graphicx}
\title{Spherical Gravitational Collapse in $4$D Einstein- Gauss- Bonnet theory }
\author[1,2]{Suresh C. Jaryal\footnote{suresh.fifthd@gmail.com}}
\author[1]{Ayan Chatterjee\footnote{ayan.theory@gmail.com}}
\affil[1]{Department of Physics and Astronomical Science, Central University of Himachal Pradesh, Dharamshala- 176215, India.}
\affil[2]{Department of Physics and Astronomical Sciences, Central University of Jammu, Samba, J\&K-(181143), India.}
\begin{document}
\date{}
\maketitle
\begin{abstract}
In this paper, we study spherical gravitational collapse
of inhomogeneous pressureless matter in a well- defined $n \rightarrow4$d  limit
of the Einstein- Gauss- Bonnet gravity.
The collapse leads to either a black hole or a massive naked singularity depending on
time of formation of trapped surfaces. More precisely, horizon 
formation and its time development is controlled by relative strengths of 
the Gauss- Bonnet coupling $(\lambda)$ 
and the Misner- Sharp mass function $F(r,t)$ of collapsing sphere. We find that,
if there is no trapped surfaces on the initial 
Cauchy hypersurface and $F(r,t)< 2\sqrt{\lambda}$\,, 
the central singularity is massive and naked. When this inequality is equalised or reversed, the 
central singularity is always censored by spacelike/timelike spherical marginally trapped surface
of topology $S^{2}\times \mathbb{R}$,
which eventually becomes null and coincides with the event horizon at equilibrium. 
These conclusions are verified for a wide class 
of mass profiles admitting different initial velocity conditions. 
Hence, our result implies that the $4$d Einstein- Gauss- Bonnet generically violates the cosmic censorship conjuncture. Further implications of this violation from the perspective of visibility of causal signals from the spacetime singularity are also discussed.
\end{abstract}
\section{Introduction}
The $n$- dimensional  Einstein- Gauss- Bonnet (EGB) theory has 
the following Lagrangian functional (we shall assume $G=1, \, c=1$):
\begin{equation}\label{total_lagrangian}
16\pi\, L[\bar{g}, \psi]=\bar{R}\,[\,\bar{g}\,]+{\lambda_{n}}\, \bar{\mathcal{G}}\,[\,\bar{g}\,]+L_{m}[\psi] -2\Lambda,
\end{equation}
where, $\bar{R}$ is the Ricci scalar, $\bar{\mathcal{G}}[\bar{g}]$ is the Gauss- Bonnet (GB)
functional in $n$- dimensions with the tensor field
$\mathcal{G}=\bar{R}^{2}-4\bar{R}_{\mu\nu}\bar{R}^{\mu\nu}+\bar{R}_{\mu\nu\lambda\sigma}\bar{R}^{\mu\nu\lambda\sigma}$,
${\lambda_{n}}$ is the $n$ dimensional GB coupling constant, $L_{m}$ is 
the Lagrangian of matter field $\psi$, 
and $\Lambda$ is the cosmological constant.
The GB Lagrangian in equation \eqref{total_lagrangian} is known to be a total derivative in 
$4$- dimensions and hence, does not contribute to the equations of 
motion \cite{Lovelock:1971yv,Lovelock:1972vz, Lanczos:1938sf,Mardones:1990qc, Torii:2008ru}.
However, it was claimed in \cite{Glavan:2019inb}
that by choosing a metric ansatz of the form:
\begin{equation}
ds^{2}=-f(r)\, dt^{2}+f(r)^{-1}\, dr^{2}+r^{2}\, d\Omega_{n-2},
\end{equation}
a rescaling of the coupling ${\lambda}\rightarrow {\lambda}/(n-4)$ before taking 
the limit $n\rightarrow 4$,
leads to a well defined theory in $4$- dimensions. This 
theory is Ostragradskii stable with $2$ dynamical gravitational degrees of freedom.
Thus, at the level of the equations of motion,
a consistent relabelling of the GB coupling constant seems to lead to
well defined set of field equations which is identical to 
$4$d limit of field equations of higher dimensional ($n\,\ge\, 5$) EGB theory.
This observation led to a surge of work to understand the possible 
physical implications of 
this $4$d EGB theory of 
gravity \cite{Konoplya:2020bxa,Wei:2020ght,Wei:2020poh,Islam:2020xmy,Ghosh:2020vpc,
Yang:2020czk,Mishra:2020gce,Doneva:2020ped,Fernandes:2022zrq}.
The interest is justified since these modified higher curvature field equations may help in 
explaining deviations from GR arising in future experiments or, may present signatures of hereto 
unknown new physics. 
So, from a purely scientific point of view, it is instructive to study 
various aspects of these equations from multiple points of view.

%

As pointed out before, the $4$d EGB theory, as proposed in \cite{Glavan:2019inb}, violates the Lovelock theorems \cite{Lovelock:1971yv,Lovelock:1972vz}. At the level of field equations, one does not have regular limit to
$4$d \cite{Gurses:2020ofy,Hennigar:2020lsl,Arrechea:2020evj}. Nevertheless, since 
 the higher curvature terms have interesting
 consequences in cosmology, astrophysics and quantum gravity, various alternate methods have been developed in which, starting from a 
 different theory, with dimensional reductions and/or field redefinitions, one is led 
 to some alternate version of a $4d$ model. 
 These attempts have led to interesting proposals \cite{Lu:2020iav, Ai:2020peo,Fernandes:2020nbq,Mahapatra:2020rds,Aoki:2020lig,Kobayashi:2020wqy}. In one of these
 suggestions, by Lu and Pang \cite{Lu:2020iav}, one begins
 with a $n$ dimensional EGB theory and carries out a Kaluza-Klein reduction to $4$d, by
 compactifying over a $(n-4)$ dimensional maximally symmetric space, and a simultaneous
 redefinition of $\lambda\rightarrow \lambda/(n-4)$.  This results in a scalar-tensor type
 theory with $3$ degrees of freedom (leaving aside the matter fields)- two gravitational, and 
 one scalar. Thus, this theory is a well defined $4$d limit of $n$- dimensional EGB theory consistent with 
 the Lovelock theorems.

 We are interested in the formation of spherically symmetric black holes from
 gravitational collapse in the Lu and Pang version of $4$d 
 EGB theory since it may illuminate questions related to the cosmic censorship hypothesis.
 To see this connection, note that the spherical black hole solution in the Lu- Pang theory is identical to 
  $n\rightarrow 4$d limit of the Boulware- Deser solution
 of $n$- dimensional EGB theory \cite{Lu:2020iav}. This
 metric in $4$d resembles the Reissner- Nordstrom spacetime (in 
 general relativity) \cite{Konoplya:2020qqh, Dadhich:2020ukj}, where the role of
 electric charge is played by square-root of the GB coupling ($\lambda$). So, 
 if $M$ is the ADM mass of the black hole, the condition $M^{2}>\lambda$ leads to a black hole solution with 
 an inner and an outer event horizon; for $M^{2}=\lambda$, the two horizons merge to form one,
 whereas for $M^{2}<\lambda$, the central singularity is globally naked.
 Thus due to this correspondence, one can propose that the nature of central
 singularity and the horizon may have dependence on relative strengths 
 of the EGB coupling constant and the (Misner- Sharp) mass inside a collapsing shell. In particular, the following behaviour should drive the gravitational collapse of spherical matter configurations leading to black hole formation:
 if the Misner- Sharp (MS) mass contained inside the collapsing spherically symmetric matter is less 
 than $\sqrt{\lambda}$, the collapse must necessarily proceed to a naked central singularity, that is, a spacetime where light is not trapped inside a horizon. On the other hand, if 
 the MS mass of collapsing shell is greater 
 than $\sqrt{\lambda}$, regular black holes with two horizons should form.
 This qualitative response and interplay of coupling constant and mass of 
 the collapsing matter in black hole formation needs to be checked 
 through a consistent and rigorous formulation of classical gravitational collapse through direct use of 
 the $4d$ Lu-Pang EGB field equations along with the notion of trapped 
surfaces. Closely related are the following questions: (a) Does singularity/ trapped region begin 
to form from collapse of the first matter shells? (b) Does the formation of central singularity depend on 
the parameters controlling the initial velocity profile and density distribution profile of 
the matter fields? (c) How does one 
distinguish naked singularity from a clothed singularity using the formulation of trapped surfaces? (d) Is 
the relation between MS mass, GB coupling and formation of trapped regions, as 
envisaged above, are in accordance with the censorship hypothesis? (e) Does 
the presence of a naked singularity violate the censorship conjecture? (f) How does 
the entire mechanism change if there is already a black hole present on 
the initial Cauchy hypersurface?  In this paper, we shall develop the theory for inhomogeneous 
gravitational collapse to answer 
 these questions.

 To understand the properties of horizons developing during the collapse process (in particular 
 the Reissner- Nordstrom like nature), we shall use 
 the formulation of trapped surfaces \cite{Penrose:1964wq}. The basic idea is as follows:
during the collapse of the matter cloud leading to a spacetime singularity, trapped surfaces should
arise inevitably. These surfaces are defined as $2$- dimensional
closed surfaces (which we shall take as round spheres) such that the expansion of the outgoing null normal to 
the sphere ($\ell^{\mu}$), and that of the outgoing null normal ($n^{\mu}$) are both negative.
Such surfaces form the trapped region, and since no light ray can extend out of this region, this
forms the black hole region.
The boundary of the trapped region is a three dimensional surface foliated by
two dimensional spheres which have vanishing outgoing expansion, $\theta_{(\ell)}=0$,
and negative ingoing expansion $\theta_{(n)}<0$. These spheres are called the marginally trapped
spheres (MTS) and the cylinder foliated by MTS is called a marginally trapped tube (MTT).
Following \cite{Ashtekar:2004cn,Ashtekar:2005ez,Andersson:2005gq,Booth:2005qc}, 
the black hole horizon, formed from the matter collapse, shall be defined as a MTT. A crucial feature 
of MTT is that they do not carry any specific signature, and hence,
can be used to infer if the horizon is growing (for which the tangent vector to MTT shall be spacelike),
of if it is in equilibrium (when 
the MTT carries null signature) \cite{Chatterjee:2021zre, Booth:2005ng,Chatterjee:2020khj}.  
Our task will be to locate
the spherical MTT during the physical process of gravitational collapse of matter, and to track 
the variations in the parameter space which affect the dynamics of MTT.
We must mention here that the marginally bound gravitational collapse model of homogeneous dust, using 
the original $4$d EGB of \cite{Glavan:2019inb},  is discussed in \cite{Malafarina:2020pvl}. However, there are several limitations in
 this work. First, the $4$d gravitational theory used in \cite{Malafarina:2020pvl} is incorrect; second, homogeneous, isotropic models cannot capture the intricacies of collapse scenarios, and third, the formulation leads to a simplistic model where the Reissner- Nordstrom like features, and the relation  of the GB coupling constant $\lambda$ towards formation of naked singularity
is not explored. Here, we consider 
the process of gravitational collapse 
of a distribution of inhomogeneous dust bounded by a spherical surface which are either
marginally bound or are bounded. This shall allow us to explore the dynamics of gravitational collapse in finer details\footnote{The $n\ge$ marginally bound collapse model of inhomogeneous dust
is discussed in \cite{Maeda:2006pm,Jhingan:2010zz}, whereas, \cite{Chatterjee:2021zre} 
discuss generic models.}.

As we shall see, the models of black hole in this theory, constructed from collapse of 
 inhomogeneous dust matter, presents a number of constraints on the
 parameters space: First, if there is no black hole on the initial hypersurface,
 the central singularity is always naked if mass function of the gravitating matter $F(r,t)$
 is less than GB coupling constant, $\lambda$. This result holds
 true for all values of coupling, but for calculations in this paper, we shall assume $\lambda <1$.
 The formation of trapped surfaces is delayed until sufficient matter is concentrated.
 Secondly, when $F(r,t)=2\sqrt{\lambda}$, a non- central marginally trapped surface
 appears covering the central singularity. Thereafter, for all $F(r,t) > 2\sqrt{\lambda}$,
 the MTT continues to cover the central singularity.
 Thirdly, this MTT may be spacelike or timelike depending
 on matter profile of the collapsing matter. However, it becomes null only when no matter
 falls through it, and eventually forms the event horizon of the black hole. 
 Fourth, the initial velocity and density profile of the collapsing matter controls the 
 time of formation of central singularity as well as the equilibriation time of the horizon.
 To exhibit the robustness of these results, we illustrate several scenarios by assuming a wide variety 
 of density profiles of collapsing matter. To ensure the correctness of our models,
 we have verified 
 that (i) our matter profiles do not admit trapped surfaces on the initial hypersurfaces (except for 
 the cases where we deliberately admit a black hole initially), and
 (ii) there are no shell crossing singularities during the process.

%
%
%
%

 The paper is organised as follows:
 In the following section, we shall present a brief discussion on dimensional reduction to $4$d Lu- Pang version of the EGB action. We shall also discuss  
 the static black hole solution (to be considered as the external spacetime) to which 
 the collapsing metric is to be 
 matched. This is followed by a discussion of the $4$d field equations
 for gravitational collapse. More specifically, using 
 inhomogeneous collapse models, also called the Lemaitre-Tolman- Bondi (LTB) models of
 pressureless dust, we derive formal equations for the collapsing matter shells, 
 the time of horizon formation, and the singularity formation time.
 This allows us to provide a complete picture of gravitational collapse leading to black holes.
 The section $3$ contains application of these equations for various matter profiles, for
 marginally bound and bounded collapse models.
 Section $4$ discusses the status of the cosmic censorship in this model. 
 The section $5$ ends with a discussion.

\section{Solutions of the Field equations}
In the approach advocated in \cite{Lu:2020iav},
one begins with a $n\ge 5$ dimensional theory, eqn. \eqref{total_lagrangian}. 
This theory is reduced over a warped metric given by:
\begin{equation}
ds_{n}^{2}=ds_{4}^{2}+\exp(\,2\phi)\, ds^{2}_{(n-4)},
\end{equation}
where the warping scalar $\phi$ depends on the $4$- dimensional spacetime coordinates.
If the compactified internal space of $(n-4)$ dimensions is flat, the reduced theory is 
given by the following $4$d Lagrangian  \cite{Lu:2020iav}:
\begin{eqnarray}\label{trace_anomaly}
L_{4}=\sqrt{-g} [R+\lambda \left\{ \phi\, \mathcal{G}+4G^{\mu\nu} \partial_{\mu}\phi\, \partial_{\nu}\phi
-4(\partial_{\mu}\phi)^{2}\, \Box \phi +2\,(\partial_{\mu}\phi)^{4}\right\}],
\end{eqnarray}
where $G_{\mu\nu}=R_{\mu\nu}-(1/2)\,R\,g_{\mu\nu}$, is the usual Einstein tensor in $4$- dimensions,
$R_{\mu\nu}, R, \mathcal{G}$ are the $4$- dimensional Ricci tensor, Ricci scalar and the Gauss- Bonnet term. 
This Lagrangian also arises in the study of spontaneously broken CFTs: 
The $4$- dimensional effective action of massless Goldstone scalar 
has terms in its effective action which are invariant 
under Diff $\times$ Weyl transformation, as well as terms arising from conformal trace anomalies. 
The $a$-trace anomaly Lagrangian is identical to eqn. \eqref{trace_anomaly} (see \cite{schwimmer} for details).

This theory has three degrees of freedom, two gravitational and one scalar. 
The scalar has no free kinetic term in this theory; it is coupled to the Einstein tensor. 
Further, the $\lambda$ dependent terms all have quartic derivatives in each term. Apart form 
the Gauss- Bonnet term, this Lagrangian belongs to the class of most general 
gravity Lagrangians, linear in the curvature scalar $R$, quadratic in $\phi$, and containing terms with quartic derivatives \cite{Amendola:1993uh,Capozziello:1999uwa,Daniel:2007kk,Sushkov:2009hk,Granda:2009fh,Gao:2010vr,Granda:2011eh}. Such theories with non- minimal derivative coupling are used to model dark matter and dark energy in cosmological spacetimes \cite{Sushkov:2009hk,Granda:2009fh,Gao:2010vr,Granda:2011eh}. However, Lagrangians with additional
Gauss- Bonnet coupling, have not been studied for gravitational collapse models, and our task in this paper
is to model this phenomenon.

The equations of motion for the metric and the scalar field can be derived from the action in 
eqn. \eqref{trace_anomaly}. They are given in detail in the Appendix 1 of this paper. The interesting equation is
the field equation of the scalar field ( given in eqn. \eqref{app_scalar_eqn} of Appendix 1). Since the kinetic term of the scalar is coupled to the Einstein tensor, this leads to a complete modification of the scalar field equations, which is now given by:
\begin{eqnarray}\label{scalar}
&&\Box\phi\,\left[(R/2)+\Box \phi +8\, (\nabla_{\mu}\phi)^{2}\right]
=R_{\mu\nu}\left[\nabla^{\mu}\phi\nabla^{\nu}\phi+\nabla^{\mu}\nabla^{\nu}\phi\right]+2(\nabla_{\mu}\nabla_{\nu}\phi)(\nabla^{\mu}\phi)(\nabla^{\nu}\phi)\nonumber\\
&&~~~~~~~~~~~~~~~~~~~~~~~~~~~~~~~~~~~~~~~~~~~~~~~~
~~+(\nabla_{\mu}\nabla_{\nu}\phi)^{2}+\mathcal{G}.
\end{eqnarray}
As we shall see below, the dynamics of the scalar field is dependent on isometries of 
the spacetimes under consideration, for static spacetime its behaviour is drastically different in comparison with dynamical spacetimes. The 
scalar equation is however very important since, the trace of the Einstein equations, and eqn.\eqref{scalar}
implies that the constraint $2R=\lambda\, \mathcal{G}$ holds on the space of solutions. This constraint implies
that on the space of solutions, the field equations of the action eqn. \eqref{trace_anomaly}, should have
similarity with those of the GR or the EGB solutions. This expectation is bourne out in both the static and dynamical solutions discussed below.

 \subsection{Static solution}
The spherically symmetric black hole solution in
this theory is also known \cite{Lu:2020iav}. The metric is given by:
\begin{equation}\label{extmetric}
ds_{4}=-f(r)\, dt^{2} +f(r)^{-1}\, dr^{2} +r^{2}\, d\Omega_{2},
\end{equation}
where $d\Omega_{2}=d\theta^{2}+\sin^{2}\theta\, d\phi^{2}$, is the $2$- dimensional round metric, and 
the metric function $f(r)$
is given by the following Boulware- Deser form \cite{Boulware:1985wk,Wheeler:1985nh,Wheeler:1985qd}:
\begin{equation}\label{metric_func}
f(r)=1+\frac{r^{2}}{2\,\lambda}\left[1-\left\{1+\frac{8\lambda M}{r^{3}}\right\}^{1/2}\, \right].
\end{equation}
The metric is a spherically symmetric, static, asymptotically flat 
black hole solution with ADM mass $M$.
The black hole has a timelike singularity at $r=0$, where the curvature scalars
blow up. Interestingly however, the metric has a nonsingular limit at that surface. Note that as
$\lambda\rightarrow 0$, the metric reduces to the Schwarzschild spacetime.
We also note that, since $\lambda >0$, there is an inner ($r_{-}$) and an outer event horizon ($r_{+}$),
located at
\begin{equation}\label{horizon_BD}
r_{\mp}=M\,\left[1\mp\left\{1-\frac{\lambda }{M^{2}}\right\}^{1/2}\right].
\end{equation}
As $\lambda=M^{2}$, the horizons merge, and we get an extremal black hole
with a single horizon, at $r_{\mp}=M$. For $\lambda>M^{2}$, there is no event horizon, and the 
singularity is globally naked. In this sense, the metric has similarity with the Reissner- Nordstrom solution,
with $\lambda$ playing the role of square of electric charge.

The role of the scalar field $\phi(r)$, which depends on the radial coordinate in 
this spacetime, is interesting. Note that the metric functions are independent of 
the $\phi(r)$ and the fall -off conditions are such that the ADM mass of the spacetime is 
not changed.
This can be understood as follows \cite{Lu:2020iav}: The field equation for $\phi(r)$, eqn. \eqref{scalar} reduces to :
\begin{equation}\label{phi_eqn}
 \left(r\,\phi^{\prime\,} -1\right)^{2} f(r) -1=0.
\end{equation}
Using $f(r)$ of eqn.\eqref{metric_func}, the regular solution of equation, \eqref{phi_eqn} is given by:
\begin{equation}
\phi(r)=\log\left[\frac{r(\cosh x -\sinh x)}{L}\right],\,\,  x(r)=\int_{r_{+}}^{r}\frac{dy}{y\sqrt{f(y)}},
\end{equation}
where $r_{+}$ is coordinate of the outer event horizon in \eqref{horizon_BD}, and $L$ is a constant of integration of eqn. \eqref{phi_eqn}, and plays the role of length scale. For the spacetimes under consideration, where compactification is carried over a flat  metric, and the black hole is spherically symmetric round metric \eqref{extmetric}, the scalar field $\phi(r)$ vanishes asymptotically. Therefore, the conditions are such that 
the ADM mass of the spacetime is not affected by the presence of the scalar.
For this specific case of spherically symmetric black hole solution, this scalar $\phi(r)$ is like background field which satisfies fixed conditions on the horizon and at asymptotic infinity. Of course, the scalar field has a completely different behaviour for other solutions of this theory, but we are not interested in them here (see  \cite{Lu:2020iav} for these solutions).
In our analysis of gravitational collapse formalism in this theory, the black hole metric of 
eqn. \eqref{extmetric} shall be taken as the external spacetime. The collapse 
of inhomogeneous matter shall be assumed to lead to this external spacetime metric for an external observer.

\subsection{Time dependent inhomogeneous metric solutions}
Let us now look into the gravitational collapse formalism for the theory in eqn. \eqref{trace_anomaly}. 
Since the scalar field has an important role to play, the first task at hand is to look into its dynamics.
The theories with scalar fields kinetically coupled to the Einstein tensor has been studied in great detail
for cosmological spacetimes \cite{Granda:2009fh,Gao:2010vr,Granda:2011eh}. It arises, quite surprisingly that, in absence of any other matter sources
or in presence of other pressureless matter, the  scalar itself behaves as a dust \cite{Gao:2010vr}. Keeping this idea in mind, we
shall consider the scalar to be a test scalar or a background spectator field, having a fixed value $\phi_{0}$, and add a dust matter 
as the energy momentum of spacetime. Then, the equations of motion, eqn. \eqref{app_Einstein} of Appendix 1, reduce to the following
\begin{eqnarray}
G_{\mu\nu}&=&8\pi T_{\mu\nu}-\lambda H_{\mu\nu}, \label{EGBQ}
\end{eqnarray}
where  $T_{\mu\nu}$ is the energy momentum tensor of dust, (the coupling $\lambda$ is in fact the quantity
$\lambda\phi_{0}$, but we shall continue to denote it as $\lambda$),
and the $H_{\mu\nu}$ is due to the Gauss- Bonnet term, which shall be taken as the effective energy momentum tensor 
having form
\begin{eqnarray}\label{hab_eqn}
H_{\mu\nu}&=&2\left[{R}R_{\mu\nu}-2R_{\mu\sigma}R^{\sigma}{}_{\nu}-2R^{\sigma\delta}R_{\mu\sigma\nu\delta}
+R_{\mu}{}^{\sigma\delta\lambda}R_{\nu \sigma\delta\lambda}\right]-\frac{1}{2}g_{\mu\nu}\, \mathcal{G}.\label{Hab}
\end{eqnarray}
The line element
of spherical symmetric space-time 
geometry can be written as 
\begin{equation}
 ds^{2}=-e^{2\Phi(r,t)}dt^2 + e^{2\Psi(r,t)}dr^2 + R(r,t)^2 d \theta^2+ R(r,t)^2 \sin^2{\theta}\, d\phi^2,
 \label{1eq1}
\end{equation}
where $\Phi(r,t)$, $\Psi(r,t)$ are the metric functions,
$R(r,t)$ is the physical radius of the matter sphere, and $\theta$, $\phi$
are the angular coordinates on the sphere.
The energy momentum tensor for pressureless dust is given by
\begin{eqnarray}
T_{\mu\nu}(r,t)=\rho(r,t)\,u_\mu u_\nu \label{tmnz}
\end{eqnarray}
where, $u^{\mu}$ is the unit time-like vector satisfying $u_\mu u^\mu=-1$,
and $\rho(r,t)$ is the energy density respectively.
In the comoving co-ordinates the four velocity is
$u^\mu=e^{-\Phi(r,t)}\,\delta^{\mu}_{0}$.
The EGB field equations which govern the dynamical development of collapsing cloud
takes the following form
 \begin{eqnarray}
F^{\,\prime}(r,t)&=&{8\pi \rho\, R^{2}(r,t)\, R^{\prime}(r,t)},\label{1eq1}\\
 \dot{F}(r,t)&=&0 \label{1eq2}\\
 \Phi(r,t)^{\prime}&=&0 \label{1eq3} \\
 \dot{R}(r,t)^{\prime}&=&\dot{R}(r,t)\,\Phi'+R^{\prime} (r,t)\,\dot{\Psi}  \label{1eq4}\\
 F(r,t)&=&R(r,t)\,(1-G+H)+ \frac{\lambda}{R(r,t)}(1-G+H)^2 \label{1eq5}
 \end{eqnarray}
where $F(r,t)$ is the Misner-Sharp mass function, and the dots, and the primes
imply differentiation with respect to $t$, and $r$ respectively.
The functions $H(r,t)$ and $G(r,t)$ are defined as $H(r,t)=\exp[-2\Phi(r, t)]\,\dot{R}(r,t){}^{2}$, and
$G(r,t)=\exp[-2\Psi(r, t)]\,R^{\prime}(r,t)^{2}$. 
Gravitational collapse is obtained by requiring that $\dot{R}<\,0.$
These equations shall lead to solutions for the metric functions
$\Phi(r,t)$, $\Psi(r,t)$, and determine the radius 
$R(r,t)$ of the matter cloud. 
First, note from eqn. \eqref{1eq2} that, $F=F(r)$, and from
eqn. \eqref{1eq3} that $\Phi(t,r)$ is a function of $t$ only. 
Naturally, one may redefine the variable $t$ to absorb the function $\Phi(t)$ in eqn. \eqref{1eq1}. 
The function $\Psi(t,r)$ similarly is obtained from eqn.\eqref{1eq4}, to 
give $\Psi(t,r)=R^{\,\prime\, 2}(r,t)/\{1-k(r)\}$. All this exercise leads to 
the following interior metric:
\begin{equation}\label{intmetric}
ds^{2}=-dt^{2}+\frac{R^{\,\prime\, 2}(r,t)}{1-k(r)}\, dr^{2} +R(r,t)^{2}\, d\Omega_{2}\,.
\end{equation}
The metric is completely determined by $R(r,t)$, and $k(r)$.
$R(r,t)$ is controlled by the density $\rho(r)$ or $F(r)$, and $k(r)$ fixes the initial energy of 
the shells, as to whether they are bounded or otherwise. 
These parameters dictate if this metric  
leads to either a black hole or a naked singularity.  
Note that this metric needs to be matched to the eqn. \eqref{extmetric}, across a $r=$constant
hypersurface. The matching of the metric and extrinsic curvatures lead to 
the condition that $F(r)=2M$, see Appendix- 3 of 
this paper for details.

%
%
\section{Horizon formation from gravitational collapse}

As we have discussed earlier, we shall use the formulation of MTT
to locate horizons. This is natural since MTT also forms the boundary of black hole region.
The MTT is located from the conditions $\theta_{(\ell)}=0$, and $\theta_{(n)}<0$. For the 
metric \eqref{1eq1}, and the choice of following null normals:
\begin{eqnarray}\label{null_normals_exp}
\ell^{\mu}&=&(\partial_{\,t})^{\mu}+ e^{-\Psi(t,r)}\,(\partial_{r})^{\mu}\\
n^{\mu}&=&(1/2)(\partial_{\, t})^{\mu} - (1/2)\,e^{-\Psi(t,r)}\,(\partial_{r})^{\mu},\label{null_normals_exp_2}
\end{eqnarray}
the eqn. \eqref{1eq5}, and the form of $\Psi(r,t)$ lead to the following expression for 
radius of MTT:
\begin{equation}\label{radius_mtt}
R(r,t)_{\scaleto{MTT}{4pt}}=\frac{F(r)}{2}\,\left[1\mp\left\{1-\frac{4\lambda }{F(r)^{2}}\right\}^{1/2}\right].
\end{equation}
This expression in eqn. \eqref{radius_mtt}
is of fundamental importance, since this value controls the time of formation of 
MTT. 
For each collapsing shell labelled by 
the coordinate $r$, the value of $R_{\scaleto{MTT}{4pt}}$ is obtained
for $F(r)=2M$, where $M$ is the total mass of the spacetime obtained from matching
the external metric in eqn. \eqref{extmetric}. So, a matching of spacetimes across $r=$constant
implies that only for $F(r)\ge 2\sqrt{\lambda}$ the is singularity covered. Until that
time for which $F(r)<2\sqrt{\lambda}$, no trapped surface forms and the singularity
remains naked.

One further advantage of the MTT formalism is that it allows a direct check of 
nature of growth of the horizon: For example, the as matter falls through, the MTT increases in radius. 
To study these changes, let $t^{\mu}=\ell^{\mu} -C\, n^{\mu}$
is tangent to the MTT. Then, $t^{\mu}$ becomes spacelike or null depending on the quantity $C$,
which is given by \cite{Chatterjee:2021zre, Booth:2006bn}:
\begin{eqnarray}
C&=&\frac{8\pi T_{\mu\nu}l^{\mu} l^{\nu}-\lambda H_{\mu\nu}l^{\mu} l^{\nu}}{4\pi/(A)-8 \pi T_{\mu\nu}l^{\mu} n^{\nu}+
\lambda H_{\mu\nu}l^{\mu} n^{\nu}},\label{OSDC}
\end{eqnarray}
where $A$ is area of the MTT. We shall see that when no matter falls in, $C=0$, and the MTT
is null, and when matter falls in, $C>0$, indicating that the MTT is spacelike. Naturally, the MTT
approaches the event horizon asymptotically when no matter falls through.
To determine the value of $C$, the geometric conditions like 
$H_{\mu\nu}l^{\mu}l^{\nu}$ and $H_{\mu\nu}l^{\mu}n^{\nu}$ 
need to be rewritten in terms of the matter variables:
\begin{eqnarray}
H_{\mu\nu}l^{\mu}l^{\nu}&=&\frac{2\rho}{R(r,t)^{2}}\left[-\frac{4F}{R}
+\frac{1}{\dot{R}(r,t)}\left\{\dot{F}(r)-\dot{R}(r,t) \right\}-\frac{\left\{F^{\prime}(r)-R^{\prime}(r,t) \right\}}{{R}^{\prime}(r,t)}\right]\nonumber\\
&&~~~~~~~~~~~~~~-\frac{6}{\dot{R}^2}\left[\left(\frac{F-R}{R}\right)^{\dot{}}\,\right]^{2}
+\frac{6}{R^{'}\,^2}\left[\left(\frac{F-R}{R}\right)^{\prime}\,\right]^{2}, \label{Hlalb}\\
H_{\mu\nu}l^{\mu}n^{\nu}&=&-\left[{\rho}^2
+\frac{7}{4}\left(\rho+\frac{4F}{R^{3}} \right)^{2}\right]+\frac{2\rho}{R}\left\{\frac{1}{\dot{R}}\left(\frac{F-R}{R}\right)^{\dot{}}
+\frac{1}{R^{'}}\left(\frac{F-R}{R}\right)^{'}\right\}.
\label{Hlanb}
\end{eqnarray}
In the following section, we shall determine the nature of the MTT using this value of $C$, 
for a wide variety of density distribution.

Two matter of vital importance which has been implemented in each of the following examples are:
(i) no crossing of shells take place during collapse, and that (ii) on the initial Cauchy hypersurface,
no black hole exists (unless we deliberately introduce them). For the first objective, 
we require that $R^{\prime}(r,t)>0$. This condition ensures physical separation of $r=$ constant
shells and, asserts that shells starting earlier in coordinate
time (or proper time in this case), shall reach the singularity earlier. The second requirement
is ensured if for the shells labelled by $r$, the no- trapping condition condition holds initially:
\begin{equation}
r> \frac{F(r)}{2}\,\left[1\mp\left\{1-\frac{4\lambda }{F(r)^{2}}\right\}^{1/2}\right].
\end{equation}
For all the matter collapse models considered below, the initial density parameters
are adjusted so that these two abovementioned conditions hold correct.  

\subsection{Marginally bound collapse}
Marginally bound collapse implies the choice $k(r)=0$, and from 
the equation (\ref{1eq5}) we obtain the equation of motion 
for the radius of the collapsing sphere $R(r,t)$. The time dependence of $R(r,t)$ 
and the time coordinates on shell are given by:
\begin{eqnarray}
\dot{R}(r,t)&=&-\frac{R(r,t)}{\sqrt{2\lambda}}\left[\sqrt{1+4\,\lambda F(r)/R(r,t)^{3}}-1\right]^{1/2} \label{dotR},\\
t_{shell}&=&\frac{2}{3}\sqrt{\lambda} \left[\left(\frac{1}{Z_0}-\arctan{(Z_0)} \right)
-\left(\frac{1}{Z}-\arctan{(Z)} \right) \right]\label{tCollapse},
\end{eqnarray}
where $Z=(1/\sqrt{2})\left[\sqrt{1+4\lambda F(r)/R^{3}}-1\right]^{1/2}$,
 and $Z_{0}=Z(R=r)$ \cite{Malafarina:2020pvl}.
The time for the formation of the singularity is obtained when the shell reaches 
$R(r,t)=0$ and is given by $t_{s}$, and the time for formation of MTT, for each shell, is obtained
from eqn. \eqref{radius_mtt}, and shall be denoted by $t_{\scaleto{MTT}{4pt}}$:
\begin{eqnarray}
t_{s}&=&\frac{2}{3}\sqrt{\lambda}\left(\frac{1}{Z_0}-\arctan{(Z_0)} \right)
-\frac{\pi}{3}\sqrt{\lambda},\label{ts}\\
t_{\scaleto{MTT}{4pt}}&=&\frac{2}{3}\sqrt{\lambda} \left[\left(\frac{1}{Z_0}-\arctan{(Z_0)} \right)
-\left(\frac{1}{Z_{1}}-\arctan{(Z_{1})} \right) \right],\label{tapp}
\end{eqnarray}
where, $Z_{1}=Z(R=R_{\scaleto{MTT}{4pt}}).$

 Let us now study how, given an initial matter density distribution, the
collapse of the matter cloud proceeds, along with the formation of MTT.
For a given initial density distribution, from the Einstein equation \eqref{1eq2},
 we can calculate the initial mass distribution of the cloud
\begin{eqnarray}
F(r)=8\pi\, \int_{0}^{r} r^{\prime\, 2}\, \rho\,(r^{\prime})\, dr^{\prime}. \label{mass1}
\end{eqnarray}
The addition of the GB term, as can be seen from eqn. \eqref{1eq5}, leads to modification 
of the dependence of mass function $F(r)$. If the mass function arising in GR
is labelled $F_{1}(r)$,  the contribution of the Gauss Bonnet term 
 and can be written as $F(r)=F_{1}(r)+\lambda F_{1}\,^2/R(r,t)^3.$
To have better understanding of this collapsing  phenomena we consider explicit examples of 
different initial density profiles. The main result of the
study is to show that the MTT begins to form only when $F(r)\ge 2\sqrt{\lambda}$, and until
$F(r)$ reaches this value, the collapse leads to globally naked massive singularities. 
For all the examples below, we take the GB coupling constant $\lambda=0.1$.

 \subsubsection*{Examples:} 
\begin{enumerate}
\item In this example, we consider the density of the cloud to be of 
the following form \cite{Booth:2005ng,Chatterjee:2020khj}:
\begin{equation}
\rho(r)=\frac{m_{0}\,\mathcal{E}(\sigma)}{r_{0}^{3}}\left[1-\erf\left\{\sigma\left(\frac{r}{r_{0}}
-1\right)\right\}\right],
\end{equation}
where $m_{0}=m(r\rightarrow \infty)$ is the total mass of the cloud,
$r_{0}$ is the location on the cloud where it matches the 
Boulware-Deser radius eqn. \eqref{horizon_BD}, and the quantity $\sigma$
controls the approach to the OSD model- larger the value of $\sigma$, closer is 
the density to uniformity. $m_{0}$ is taken to be $1$. The function $\mathcal{E}(\sigma)$
has the following form:
\begin{equation}
\mathcal{E}(\sigma)=3\sigma^{3}\, \left[2\pi\sigma(2\sigma^{2}+3)(1+\erf \sigma)
+4\sqrt{\pi}\exp(-\sigma^{2})(1+\sigma^{2})\right]^{-1},
\end{equation}
and $\erf$ is the usual error function.
We consider the cases where $\sigma=5$ and $15$. The graphs
are given in fig. \eqref{fig:OSDKg0_example1}. The graphs show that 
as the shells begin to collapse, the central singularity is formed but is not shrouded
by MTT. The formation of trapped spheres are delayed.
The MTT begins only after sufficient number of shells have fallen in. For example
in the fig. \eqref{fig:OSDKg0_example1}d, only after the shell labelled by $R=1.36$
fallen in, that the MTT begins. In fact, the MTT is noncentral and begins only after the mass 
content of the singularity is $\ge 0.3$. Before that, the spacetime has no marginally trapped surfaces.
Note that until the MTT forms, the singularity is massive and globally naked: the 
interior metric is matched with the external metric in eqn. \eqref{extmetric} admitting naked 
central singularity for each shell radius.   
The graph in fig. \eqref{fig:OSDKg0_example1}b shows that
the MTT has $C>0$, meaning that the MTT is spacelike and becomes null
only when no more matter falls through, after the shell labelled by $r=3$. 
The MTT at equilibrium is at $r=1.95$, which is exactly the Boulware- Deser event horizon. 
 
\begin{figure}[h]
	\begin{subfigure}{.45\textwidth}
		\centering
		\includegraphics[width=\linewidth]{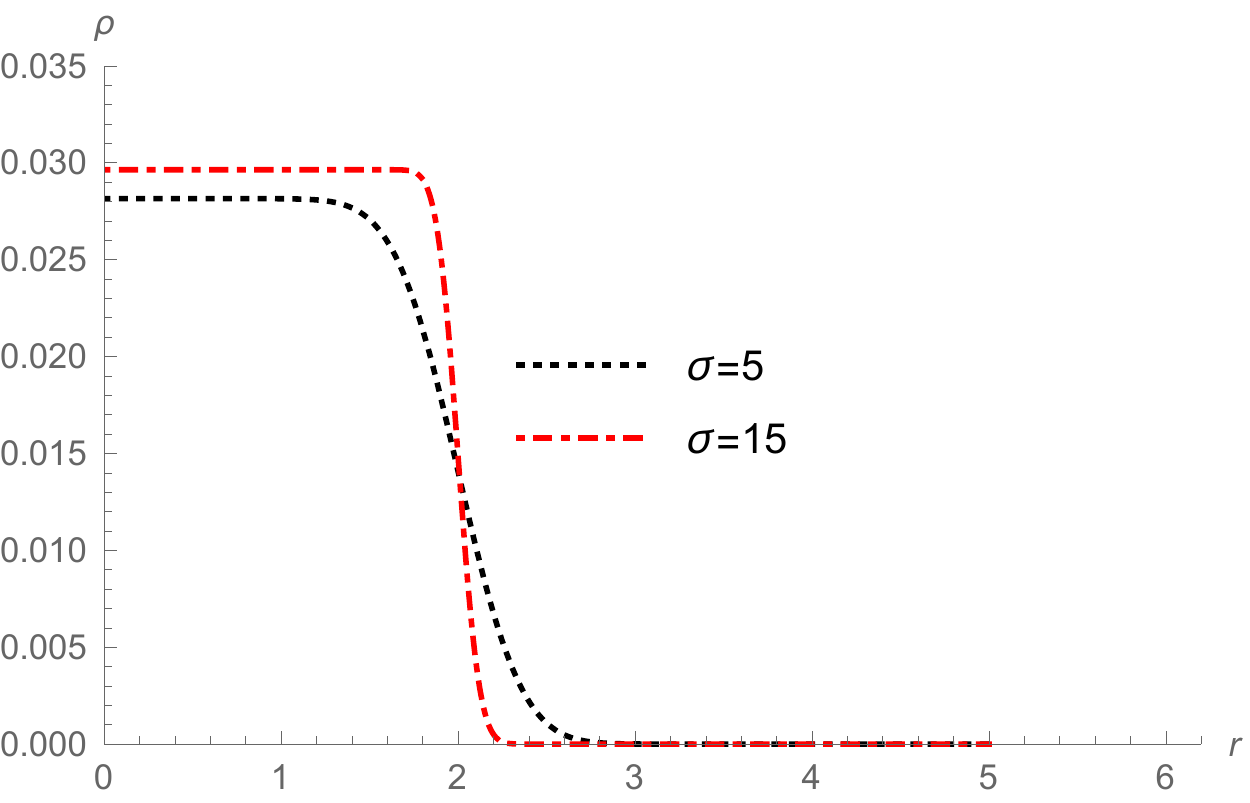}
		\caption{}
	\end{subfigure}
	\begin{subfigure}{.5\textwidth}
		\centering
		\includegraphics[width=\linewidth]{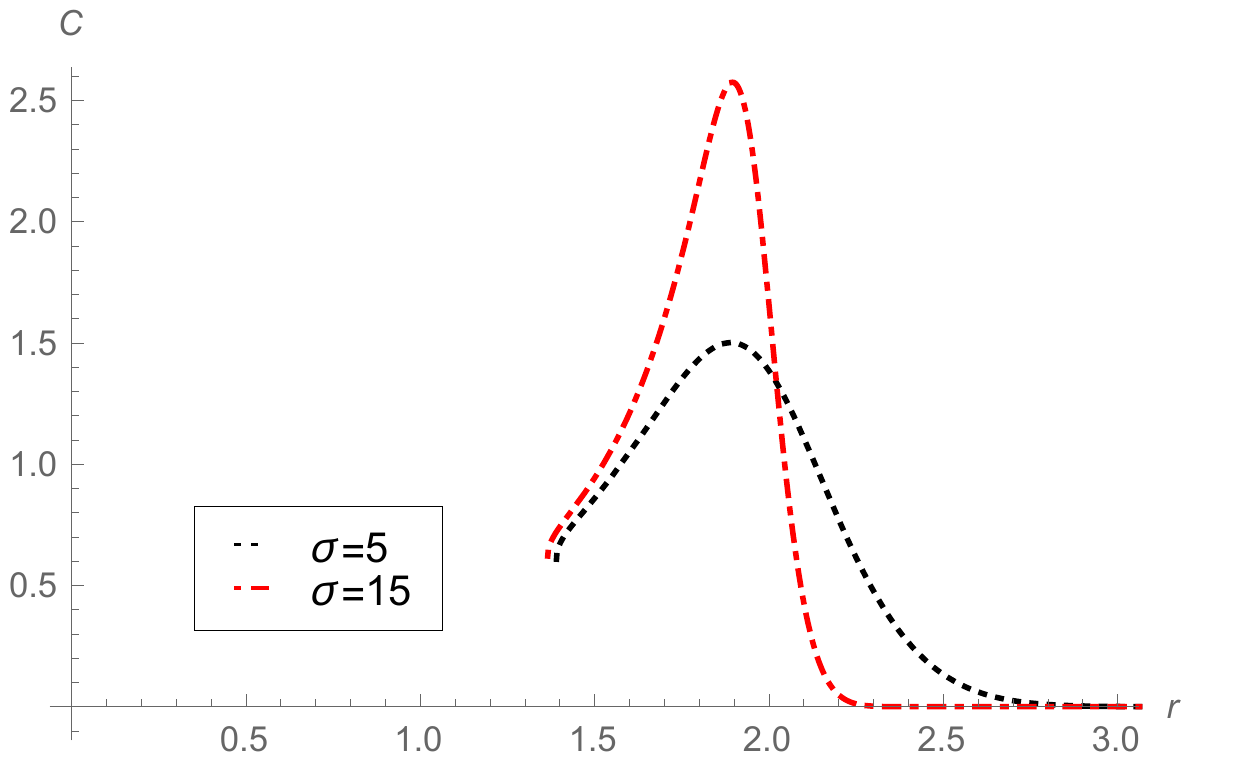}
		\caption{}
	\end{subfigure}
\\
	\begin{subfigure}{.45\textwidth}
		\centering
		\includegraphics[width=\linewidth]{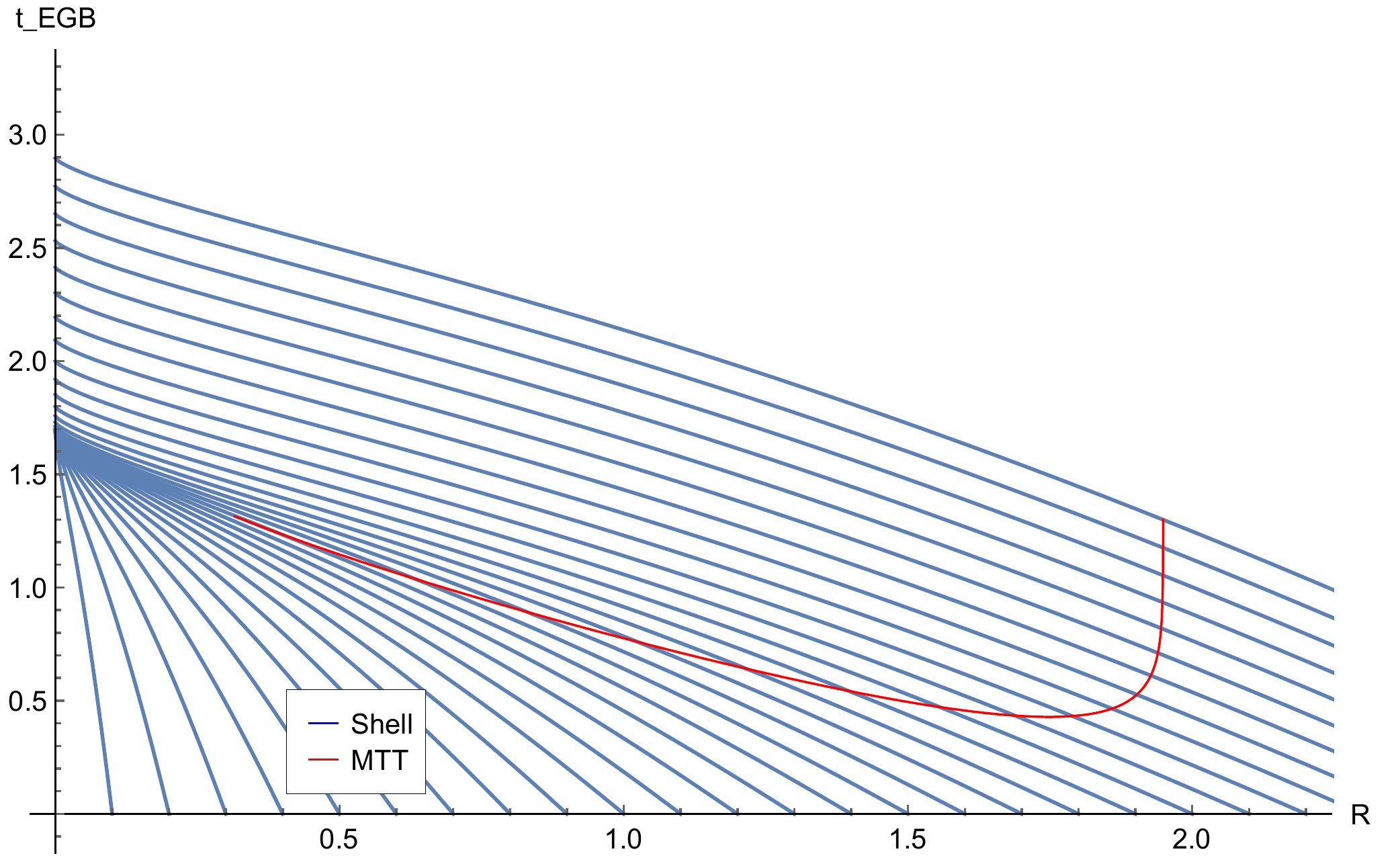}
		\caption{}
	\end{subfigure}
   \begin{subfigure}{.45\textwidth}
		\centering
		\includegraphics[width=\linewidth]{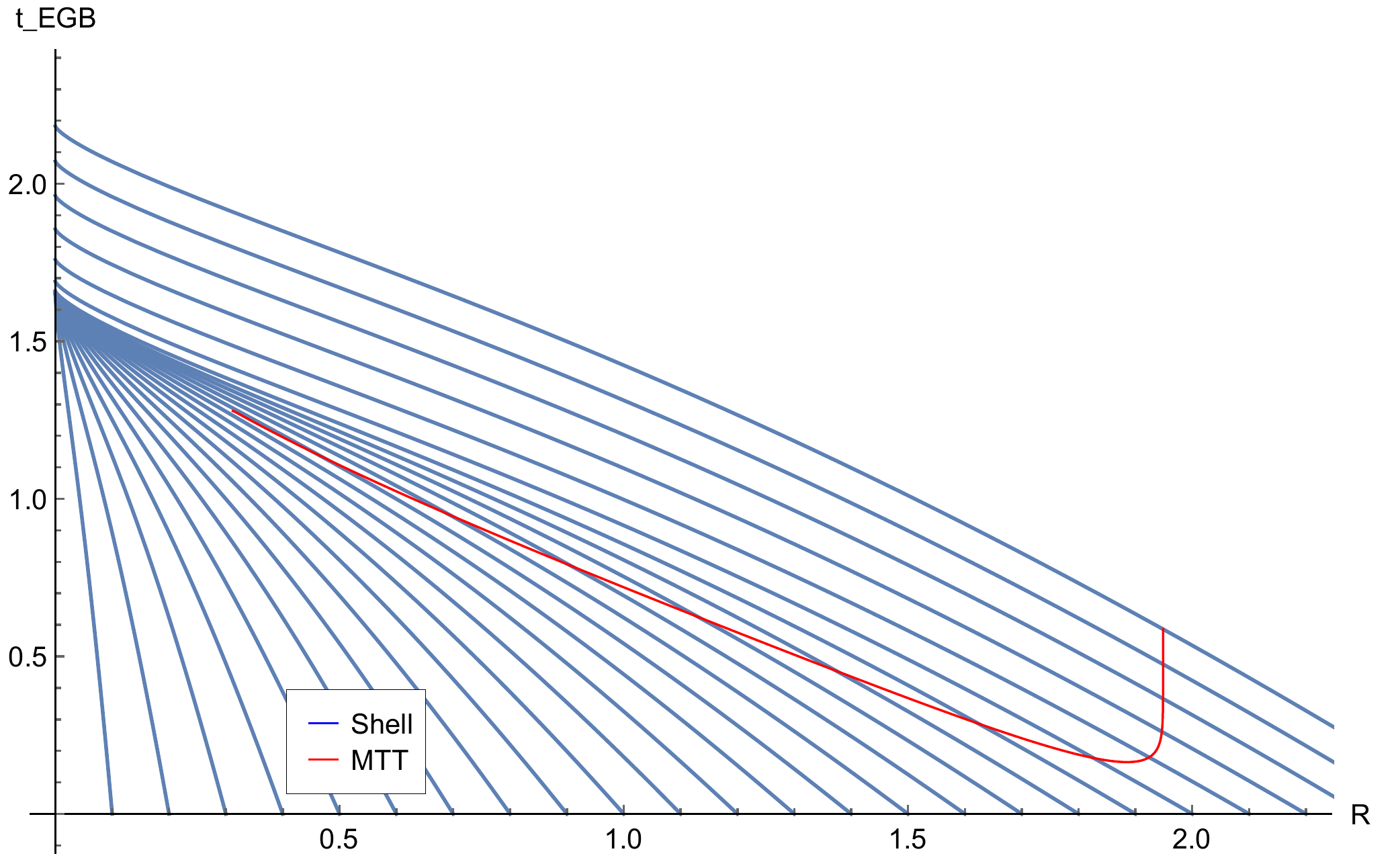}
		\caption{}
	\end{subfigure}
	\caption{This gives the gravitational collapse for the OSD profile discussed above where (a)gives 
	the density fall-off, (b) gives the values for $C$ and hence the signature of MTT, (c) gives 
	the formation of MTT for $\sigma=5$, and
	$\lambda=0.1$. Here, the MTT begins to form only after 
	the shall labelled by $R=1.39$ falls in. Until then,
	the singularity is naked and is matched with 
	the external metric in eqn.\eqref{extmetric} admitting naked 
	central singularity. Similarly, The graph (d) shows the MTT for $\sigma=15$ along with 
	the shell coordinates for $\lambda=0.1$. Here, the MTT begins only after the shall labelled by $R=1.36$
	falls in. The straight lines of MTT in (c) and (d) represents 
	the isolated horizon phase obtained at $R=1.95$, and when the horizon is null.
	The MTT looks like a timelike surface in (c), and (d), but that is only due to the choice of foliation.}
	\label{fig:OSDKg0_example1}
\end{figure}
%
\item Let us consider the case where the initial mass density of the collapsing matter is of 
the following form: 
\begin{equation}
\rho_{1}(r)=(3M/2500)(10-r)\,\Theta(10-r),
\end{equation}
where $M=1$. For this density profile too, initially, the central singularity is naked and massive.
The trapped surfaces form only after some massive shells fall in. As can be seen from the graph
in fig. \eqref{fig:ltbkg0_example2}, the MTT begins after shell $R(r,t)=5$ has collapsed, and the mass
content of the singularity $\ge 0.3$. The MTT
becomes null and approaches the isolated phase, after the shell at $R(r,t)=10$. The graph
in fig.\eqref{fig:ltbkg0_example2}a shows that after this,
no more mass falls in.
%
\begin{figure}[h!t]
	\begin{subfigure}{.55\textwidth}
		\centering
		\includegraphics[width=\linewidth]{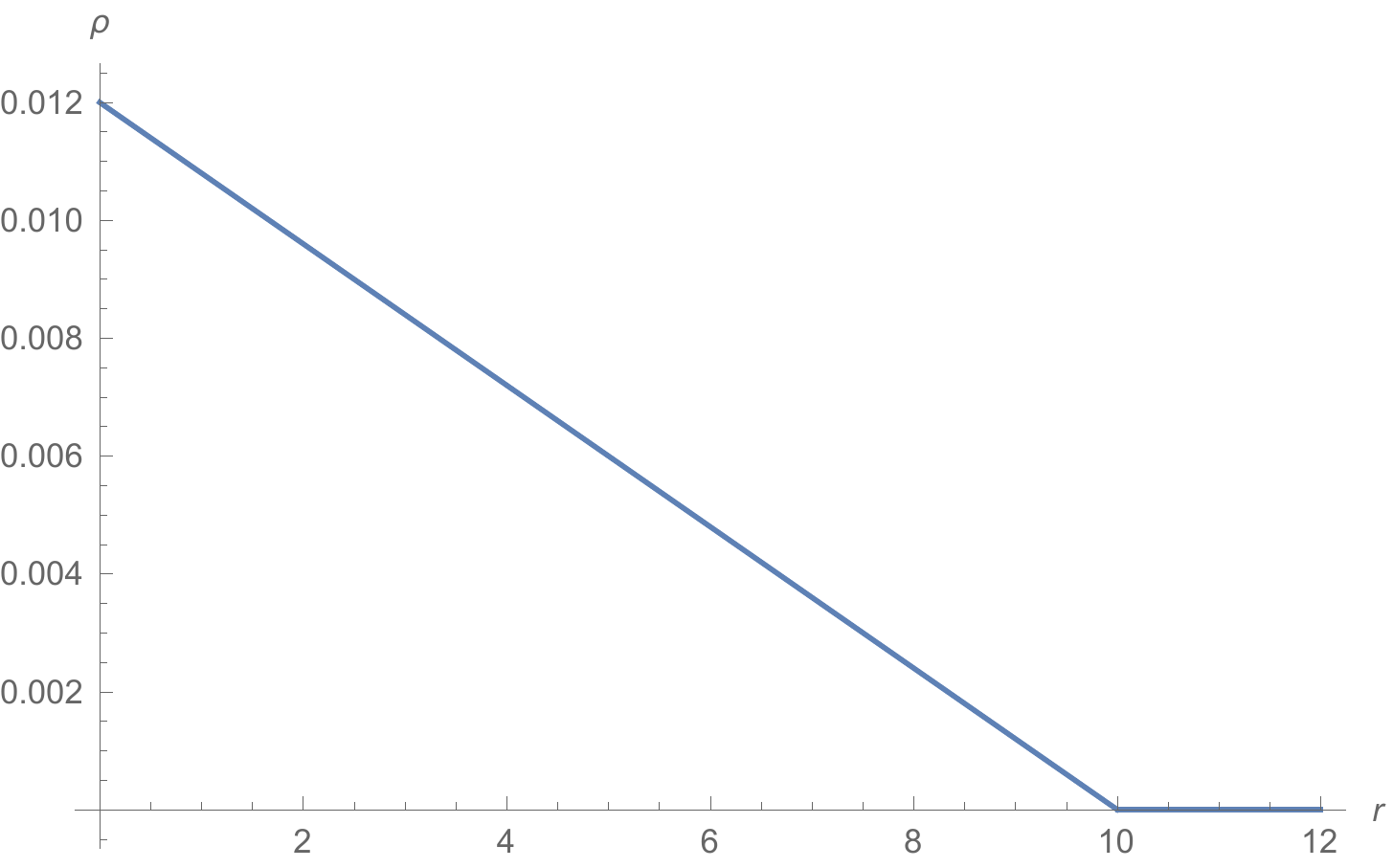}
		\caption{}
	\end{subfigure}
	\begin{subfigure}{.45\textwidth}
		\centering
		\includegraphics[width=\linewidth]{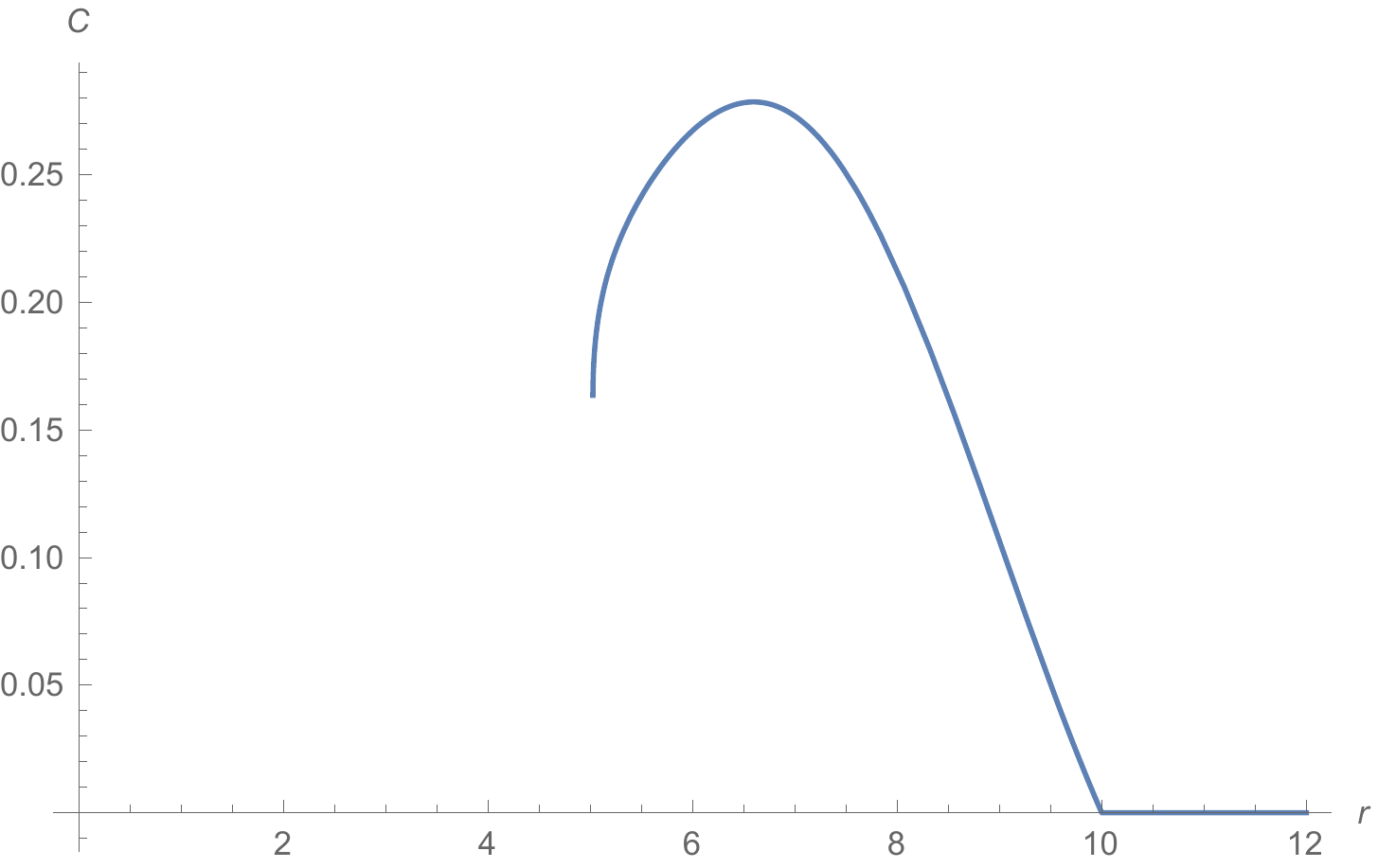}
		\caption{}
	\end{subfigure}
	\begin{subfigure}{.45\textwidth}
		\centering
		\includegraphics[width=\linewidth]{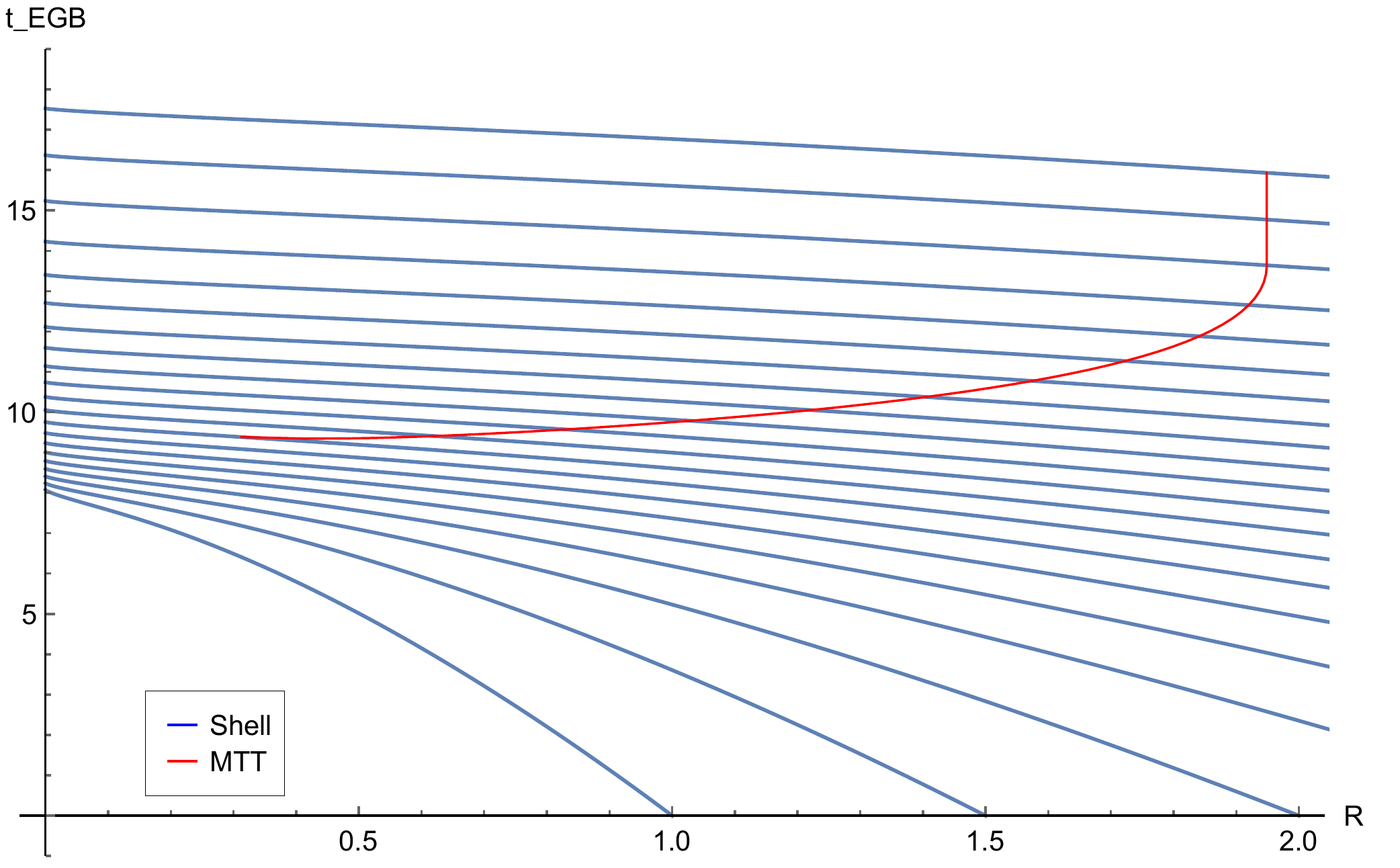}
		\caption{}
	\end{subfigure}
	\caption{The graphs show the (a) density distribution, (b) values of $C$ which dictates 
	the nature of MTT,
		and (c) formation of MTT along with the shells for $\lambda=0.1$. The MTT begins 
		non-centrally, and only when the mass content of the singularity is $\ge$ 0.3.
		The straight lines of MTT in (c), after approximately $r=10$, represents the isolated horizon phase when MTT becomes null.}
	\label{fig:ltbkg0_example2}
\end{figure}
%

\item Let us consider another density profile with the following form:
\begin{equation}
\rho(r)=\frac{m_{0}}{8\pi r_{0}^{3}}\exp(-r/r_{0}),
\end{equation}
where $m_{0}=1$ is the total mass of the matter cloud, $r_{0}$ is a parameter which indicates 
the distance where the density of the cloud decreases to $[\rho\,(0)/e]$. As 
seen from fig. \eqref{fig:ltbkg0_example3}c, the MTT is non- central, and  
begins at approximately with the shell $R(r,t)=20$, when mass of the singularity $\ge 0.3$.
After this shell has collapsed, MTT becomes spacelike (as seen from $C> 0$) and remains so 
throughout the collapse until it reaches the isolated phase, when $C=0$, 
see fig.\eqref{fig:ltbkg0_example3}b. 
The isolated phase at approximately $R(r,t)=100$, after which all shells are massless.
%
\begin{figure}[h!]
	\begin{subfigure}{.55\textwidth}
		\centering
		\includegraphics[width=\linewidth]{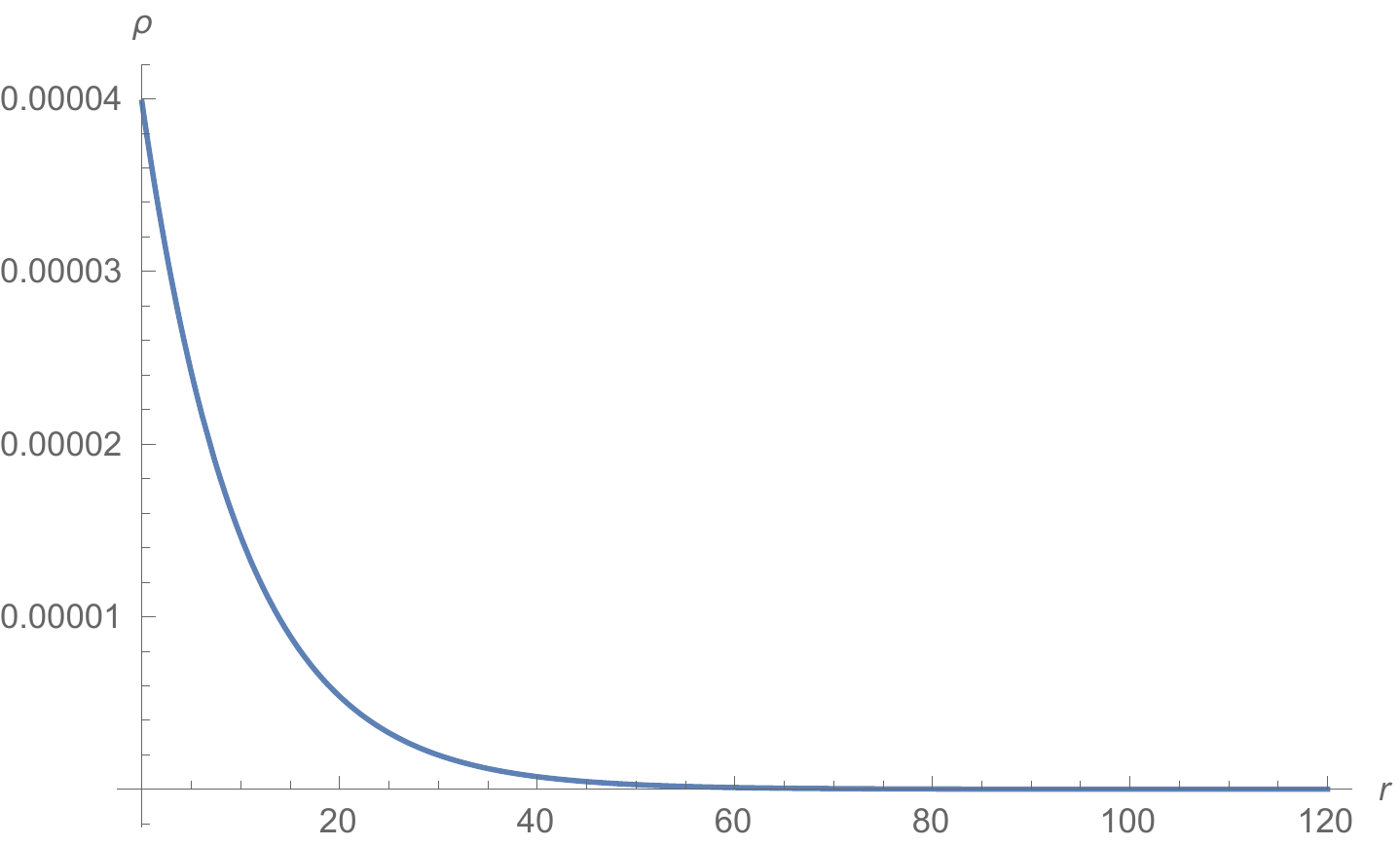}
		\caption{}
	\end{subfigure}
	\begin{subfigure}{.45\textwidth}
		\centering
		\includegraphics[width=\linewidth]{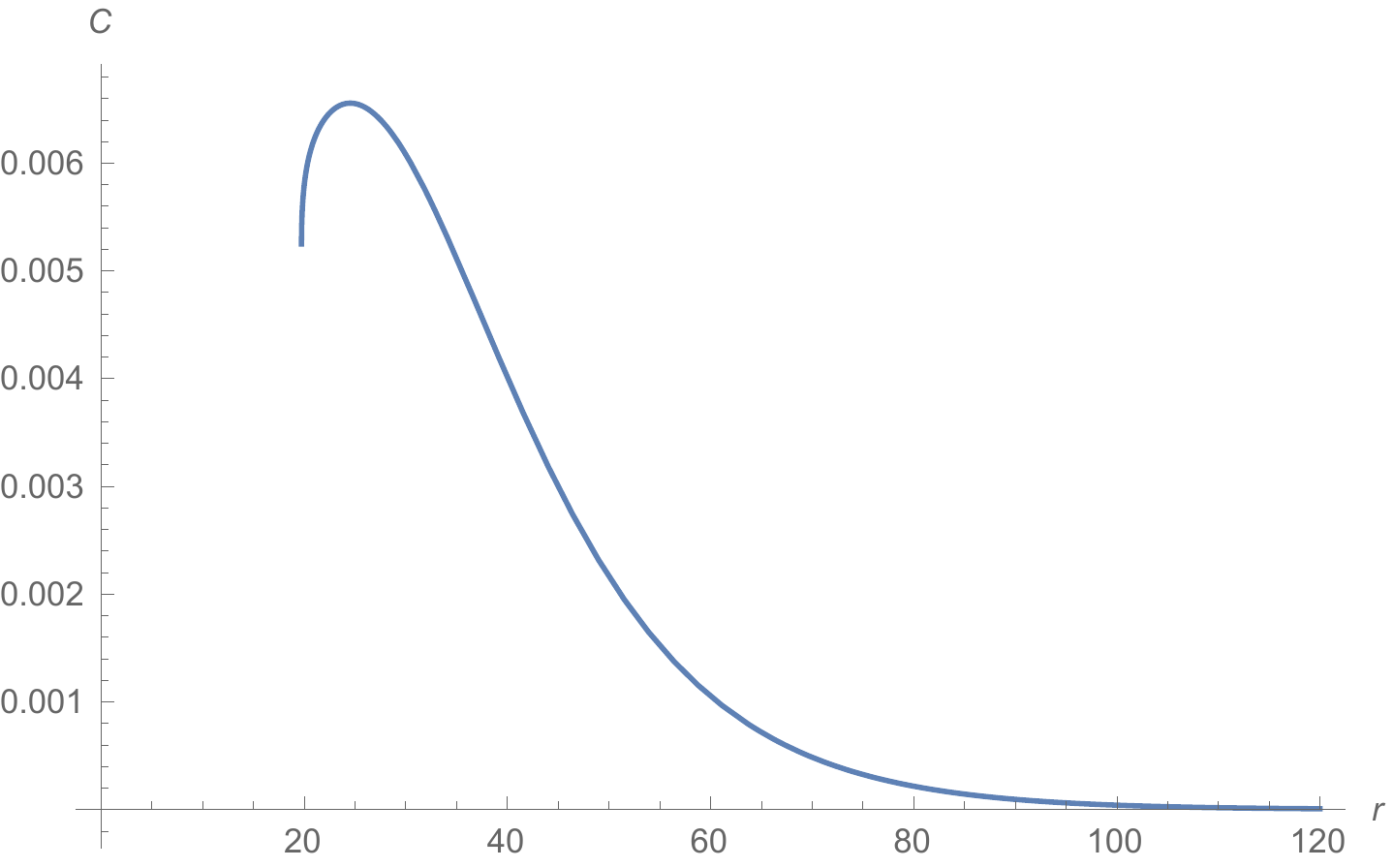}
		\caption{}
	\end{subfigure}
	\begin{subfigure}{.55\textwidth}
		\includegraphics[width=\linewidth]{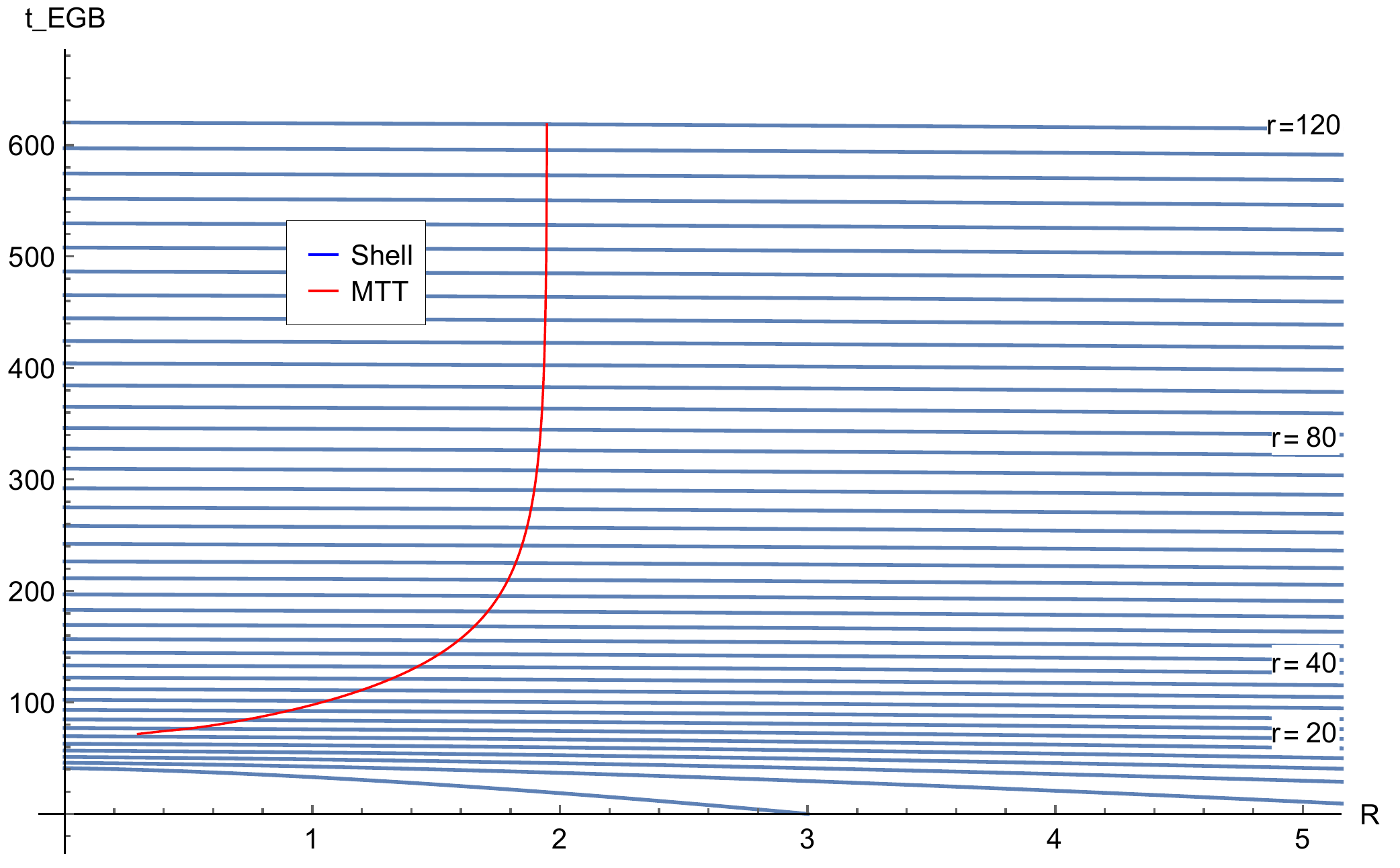}
		\caption{}
	\end{subfigure}
	\caption{The graphs show the (a) density distribution, (b) values of $C$,
		and (c) formation of MTT along with the shells for $\lambda=0.1$. The MTT is non-central and spacelike for $20<R(r,t)<100$.
		The straight lines of MTT in (c) represents the isolated horizon phase.}
	\label{fig:ltbkg0_example3}
\end{figure}
\item Here, the situation is as follows. A black hole exists on the initial hypersurface, and on it, a 
large shell of constant density falls. The nature of MTT here is drastically 
different in nature from that described in the previous examples. 
Here, we shall observe formation of timelike MTTs. The initial density distribution of 
the collapsing matter is \cite{Booth:2005ng,Chatterjee:2020khj}
\begin{equation}
\rho(r)=\frac{3\,m_{0}\left[\erf\left(\frac{r-r_{1}}{M} \right)-\erf\left(\frac{r-r_{2}}{M} \right) \right]}{4\pi \left(r_{2}-r_{1}\right) \left(2 r_{1}^2+2 r_{2}^2+2 r_{1}r_{2}+3 M^{2}\right)},
\end{equation}
where $m_{0}=600M$, $r_{1}=100M$, $r_{2}=2000M$. The form of the density is given in fig.\eqref{fig:ltbkg0_example4}a. If we assume that
the initial black hole has the Schwarzschild radius given by $\bar{r}=2M$, then the mass
for each shell of radius $r (r> \bar{r})$ is then $m(r)=M+4\pi\, \int_{\bar{r}}^{r}\rho(\hat{r})
\hat{r}^{2}\, d\hat{r}$.
\begin{figure}[h!]
	\begin{subfigure}{.55\textwidth}
		\centering
		\includegraphics[width=\linewidth]{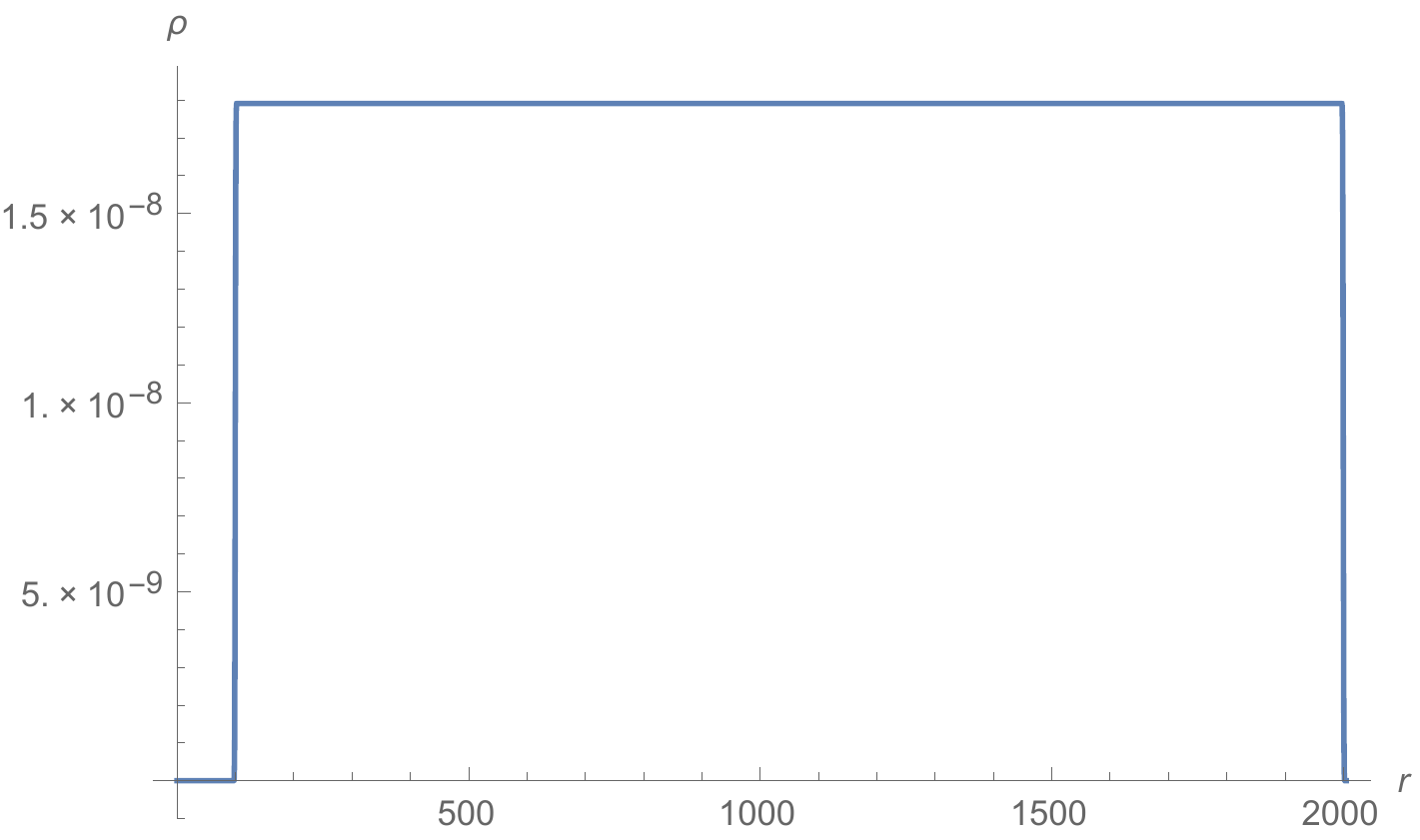}
		\caption{}
	\end{subfigure}
	\begin{subfigure}{.45\textwidth}
		\centering
		\includegraphics[width=\linewidth]{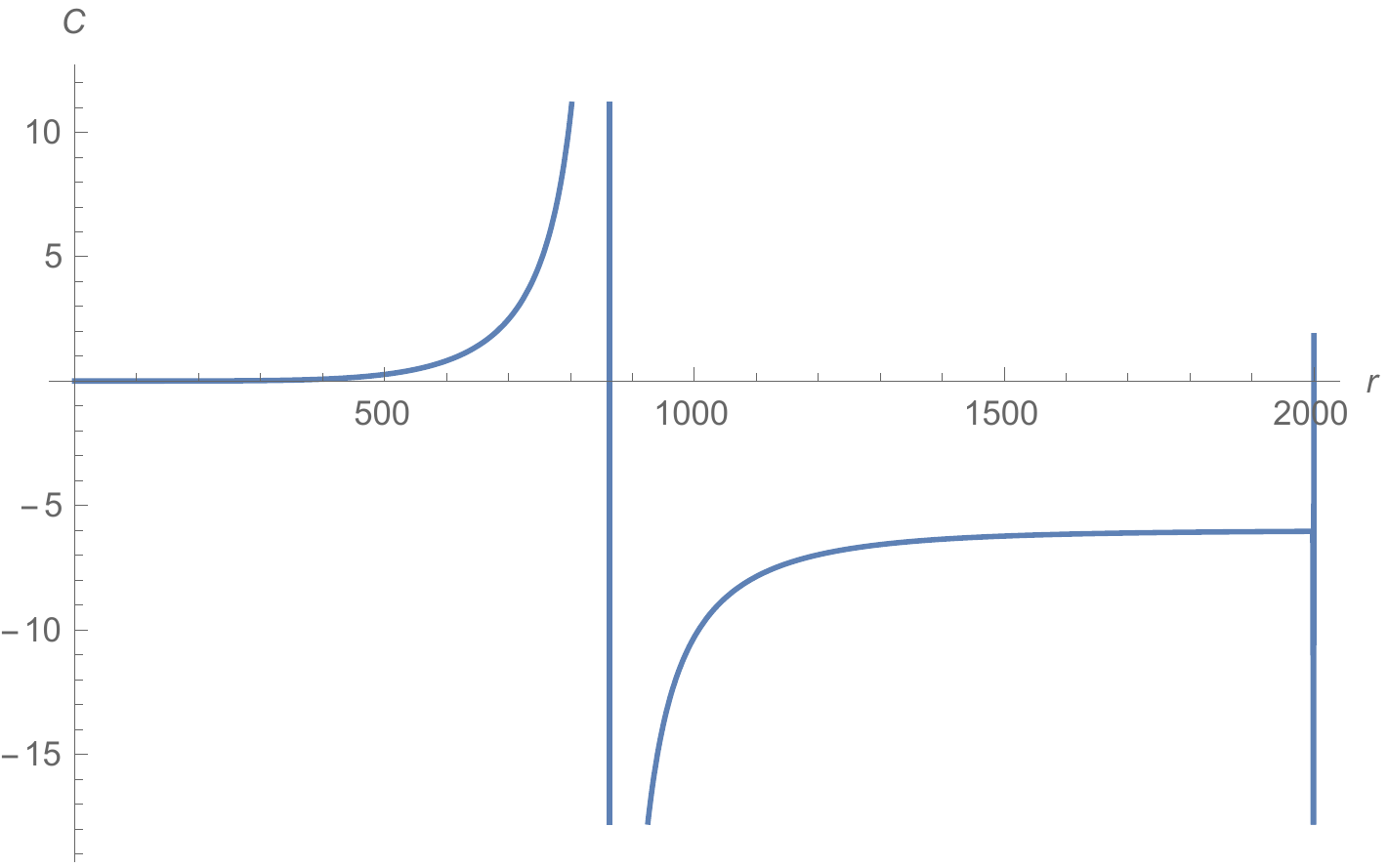}
		\caption{}
	\end{subfigure}
	\begin{subfigure}{.55\textwidth}
		\centering
		\includegraphics[width=\linewidth]{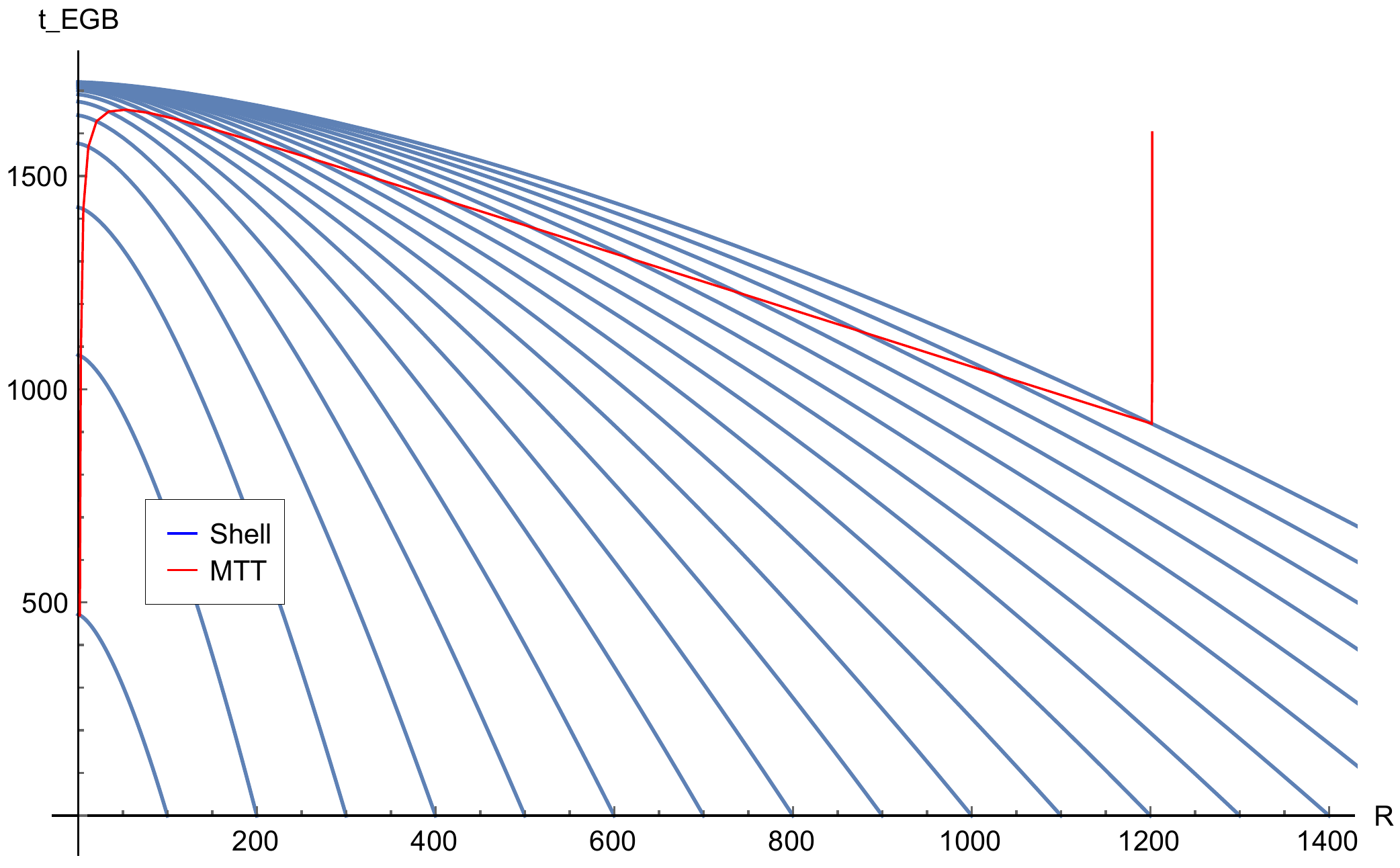}
		\caption{}
	\end{subfigure}
	\caption{The graphs show the (a) density of the collapsing shells,
		(b) values of $C$ and (c) formation of MTT along with the shells  for $\lambda=0.1$. The MTT is timelike. It first forms at $t=900$ approximately, for the shell $r=2000$, when radius of the MTT
		is approximately $R_{\scaleto{MTT}{4pt}}=1200$. 
		The MTT bifurcates in one direction to reach the isolated phase 
		and in another direction to match with the
		evolving horizon from the previous black hole. Note that here, 
		the central singularity is always covered. }
	\label{fig:ltbkg0_example4}
\end{figure}
%
The MTT forms at two places, (i) at the center of the collapsing sphere, at $t=500$, when 
the shell labelled by $r=100$ collapses, and (ii) at approximately $t=1000$, with the collapse of the 
last shell at $r=2000$. The outer MTT 
then bifurcates, and proceeds as a timelike membrane(as can be seen from the values  of $C$) to meet 
the inner MTT. The other part of the outer MTT becomes null and reaches equilibrium.
 The timelike/null nature of the MTT is also confirmed from 
 the values of $C$ in fig.\eqref{fig:ltbkg0_example4}b. Note that since a black hole exists initially,
 the MTT begins centrally with the fall of the first shell, and the singularity is censored.
\item Let us consider another density profile with the following form:
\begin{equation}
\rho(r)=\frac{\alpha\mu}{2\pi^{2}\,r_{0}\,r^{2}}\, \sin^{2}(\alpha r/r_{0}),
\end{equation}
where $r_{0}=1$ is a length scale, $\mu= (8\pi\, r_{0}/5)$, is the measure of mass
between successive shells, and $\alpha=1/10$.  As 
seen from fig.\eqref{spacelike_shell}c, the MTT is non- central, and  
begins at approximately with the shell $R(r,t)=20$, when mass of the singularity $\ge 0.3$.
Now, just as the density of matter approaches zero, the MTT tends to become null.
This behavior is clear from the graphs of MTT in fig.\eqref{spacelike_shell}c, 
and the values of C in fig.\eqref{spacelike_shell}b.
The isolated phase continues to appear in between the spacelike phase, throughout the process.

%
\begin{figure}[h!]
	\begin{subfigure}{.55\textwidth}
		\centering
		\includegraphics[width=\linewidth]{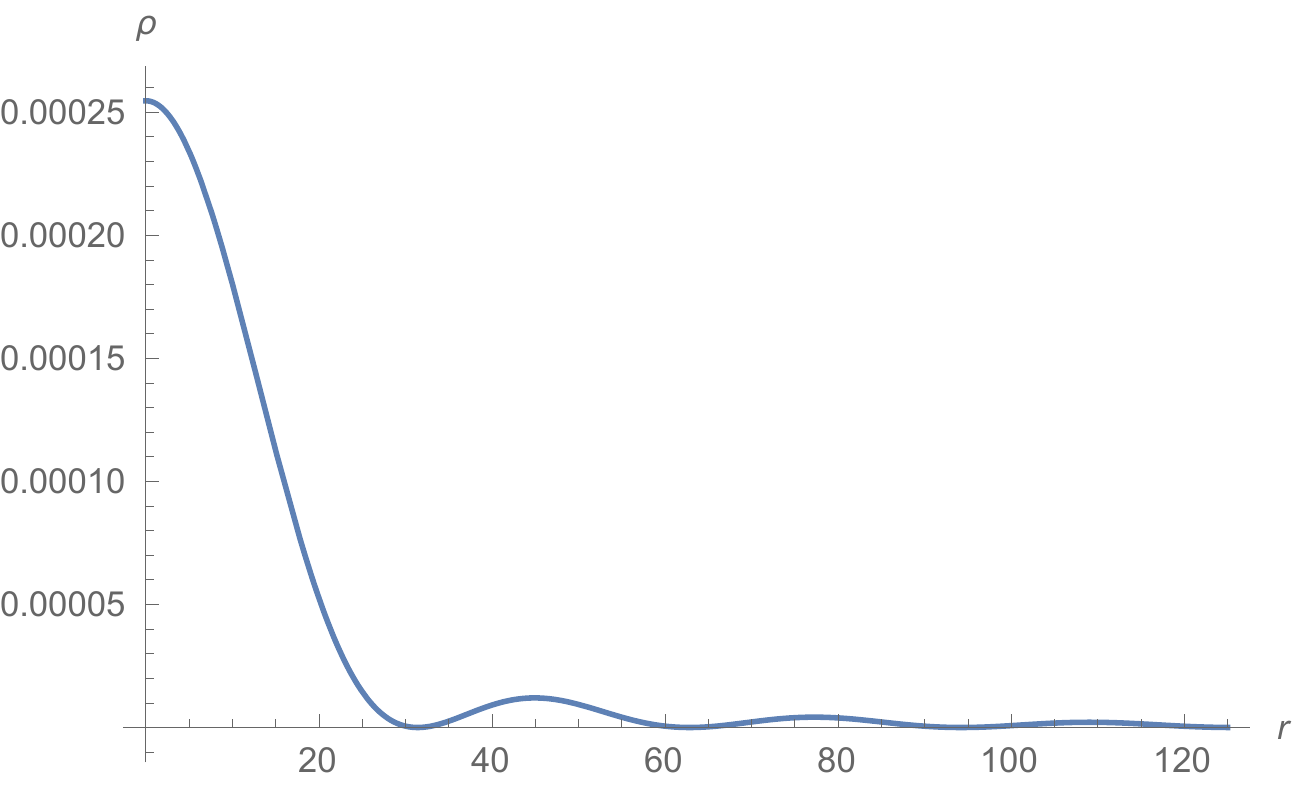}
		\caption{}
	\end{subfigure}
	\begin{subfigure}{.45\textwidth}
		\centering
		\includegraphics[width=\linewidth]{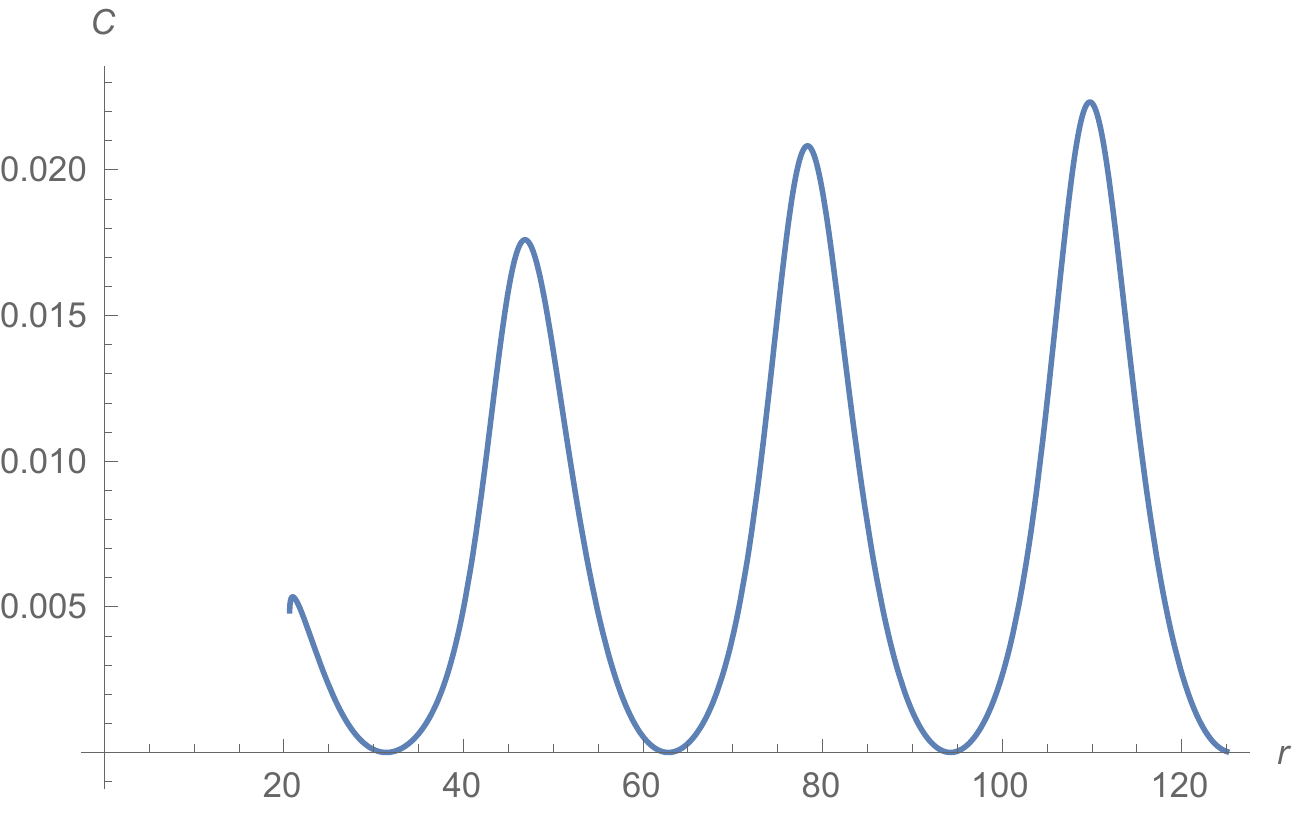}
		\caption{}
	\end{subfigure}
	\begin{subfigure}{.55\textwidth}
		\includegraphics[width=\linewidth]{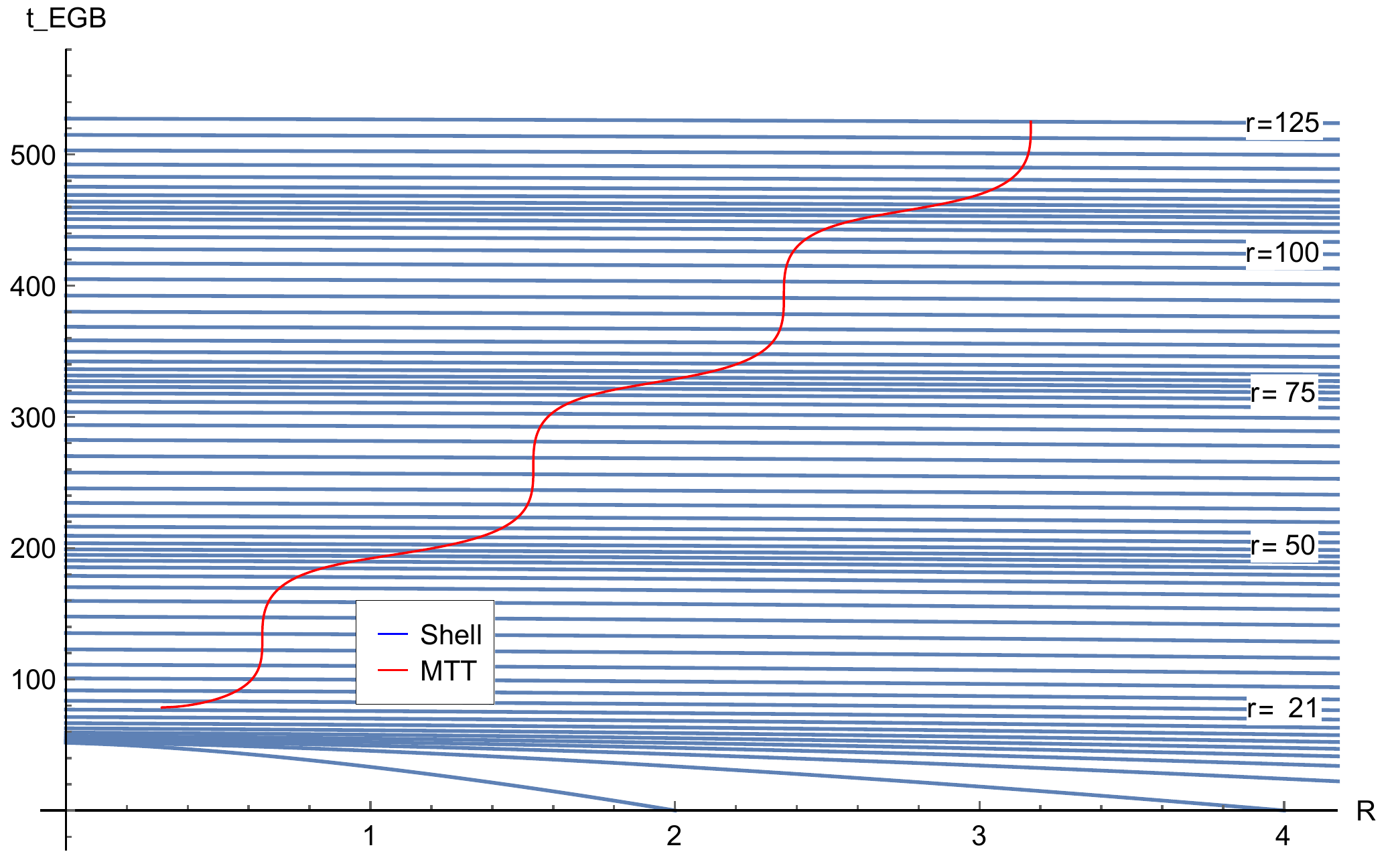}
		\caption{}
	\end{subfigure}
	\caption{The graphs show the (a) density distribution, (b) values of $C$,
		and (c) formation of MTT along with the shells for $\lambda=0.1$. The MTT is oscillates between null and spacelike nature, in accordance with the fall of matter.}
	\label{spacelike_shell}
\end{figure}

\end{enumerate}
\subsection{Bounded collapse}
For the bounded collapse, we shall consider the choice of the function $k(r)<0$.
Just like the case for marginally bounded collapse, the dynamics the shell radius and the time coordinate
on the shell can be easily obtained from the equations of motion: 
\begin{eqnarray}
\dot{R}(r,t)&=&-\frac{R}{\sqrt{2\lambda}}\left[{\sqrt{1+\frac{4\,\lambda F}{R^3}}-\left(1+\frac{2\lambda k}{R^2}\right)}\right]^{1/2}. \label{dotRk}
\end{eqnarray}
To solve this equation, we assume $k(r)=F(r)/r$, and define a new variable
\begin{equation}
x=-R(r,t)^{2}-2\,\lambda\,k+\sqrt{R(r,t)^{4}+4\,\lambda\,F(r)\,R(r,t)},
\end{equation}
Naturally, one defines 
the associated quantities: $x_{0}=-r^2-2\,\lambda\,k+\sqrt{r^{4}+4\,\lambda\,F\,r}$, and 
$x_{\scaleto{MTT}{3pt}}=-R_{\scaleto{MTT}{3pt}}^2-2\,\lambda\,k+\sqrt{R_{\scaleto{MTT}{3pt}}^{4}+4\,\lambda\,F\,R_{\scaleto{MTT}{3pt}}}$. Then, the time of formation of the
singularity and the MTT is given by a long expression, and is given in the appendix.
We shall use these expressions to study MTT formation.

 \subsubsection*{Examples:} 
(i) Let us consider a Gaussian profile with density given by 
the following form \cite{Booth:2005ng,Chatterjee:2020khj}
:
\begin{equation}
\rho(r)=\frac{m_{0}}{\pi^{3/2}r_{0}^{3}}\exp (-r^{2}/r_{0}^{2}),
\end{equation}
where $m_{0}=1$ is the total mass of the matter cloud, $r_{0}$ is a parameter which indicates 
the distance where the density of the cloud decreases to $[\rho\,(0)/e]$. 
%
\begin{figure}[h]
	\begin{subfigure}{.45\textwidth}
		\centering
		\includegraphics[width=\linewidth]{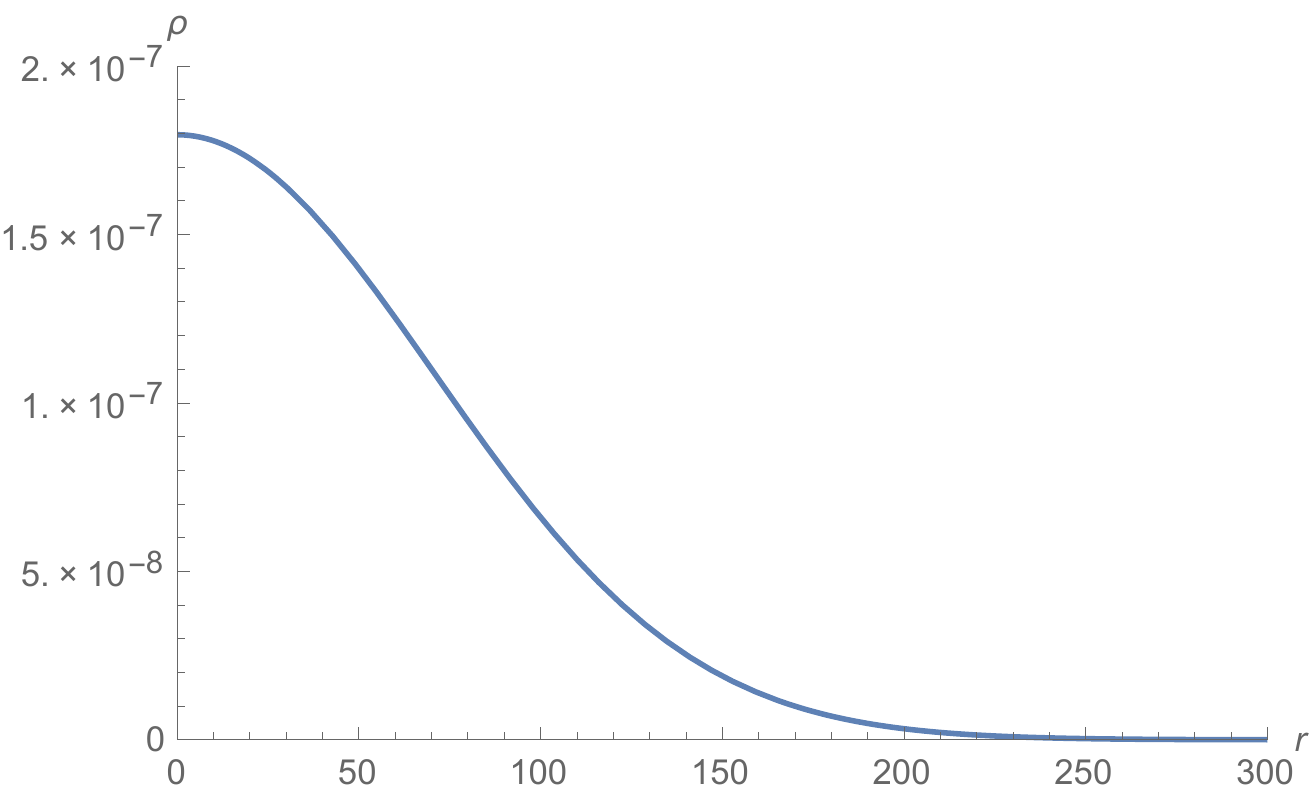}
		\caption{}
	\end{subfigure}
	\begin{subfigure}{.45\textwidth}
		\centering
		\includegraphics[width=\linewidth]{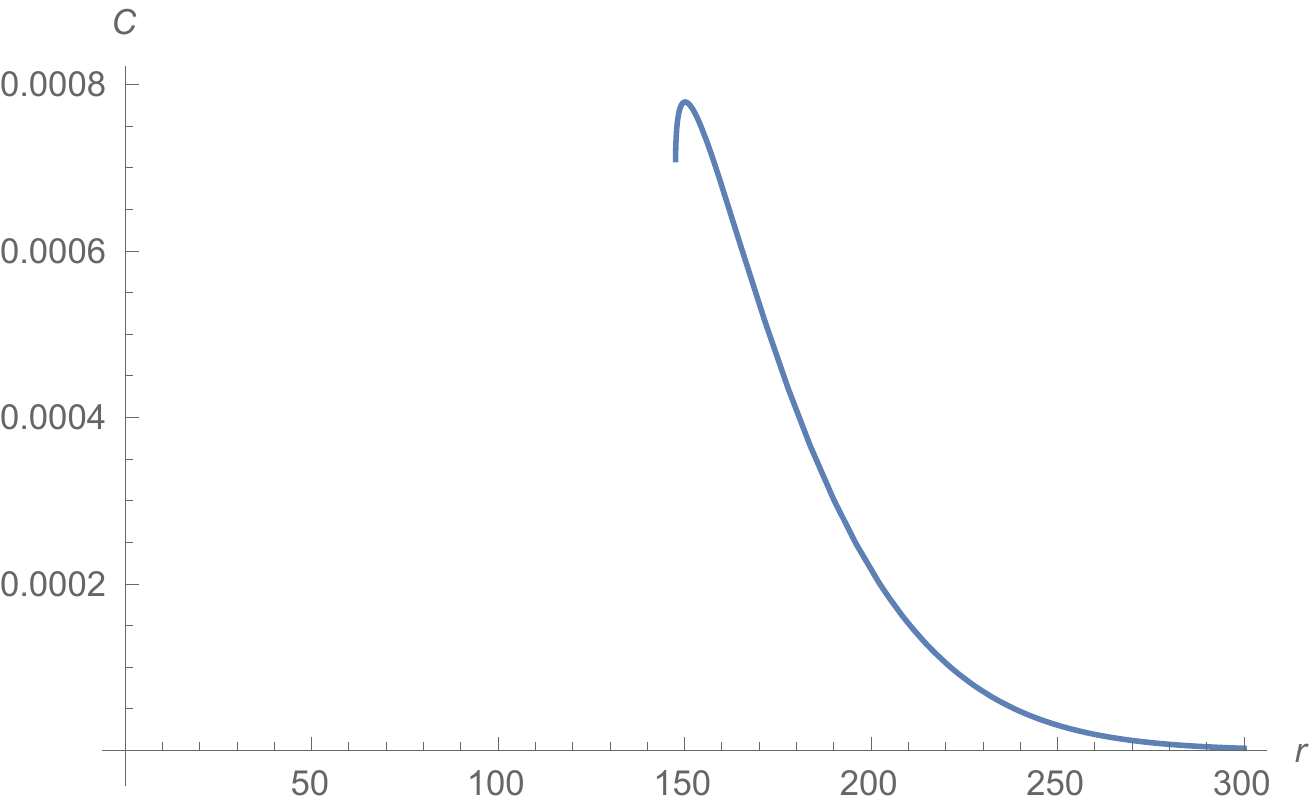}
		\caption{}
	\end{subfigure}
	\begin{subfigure}{.55\textwidth}
		\includegraphics[width=\linewidth]{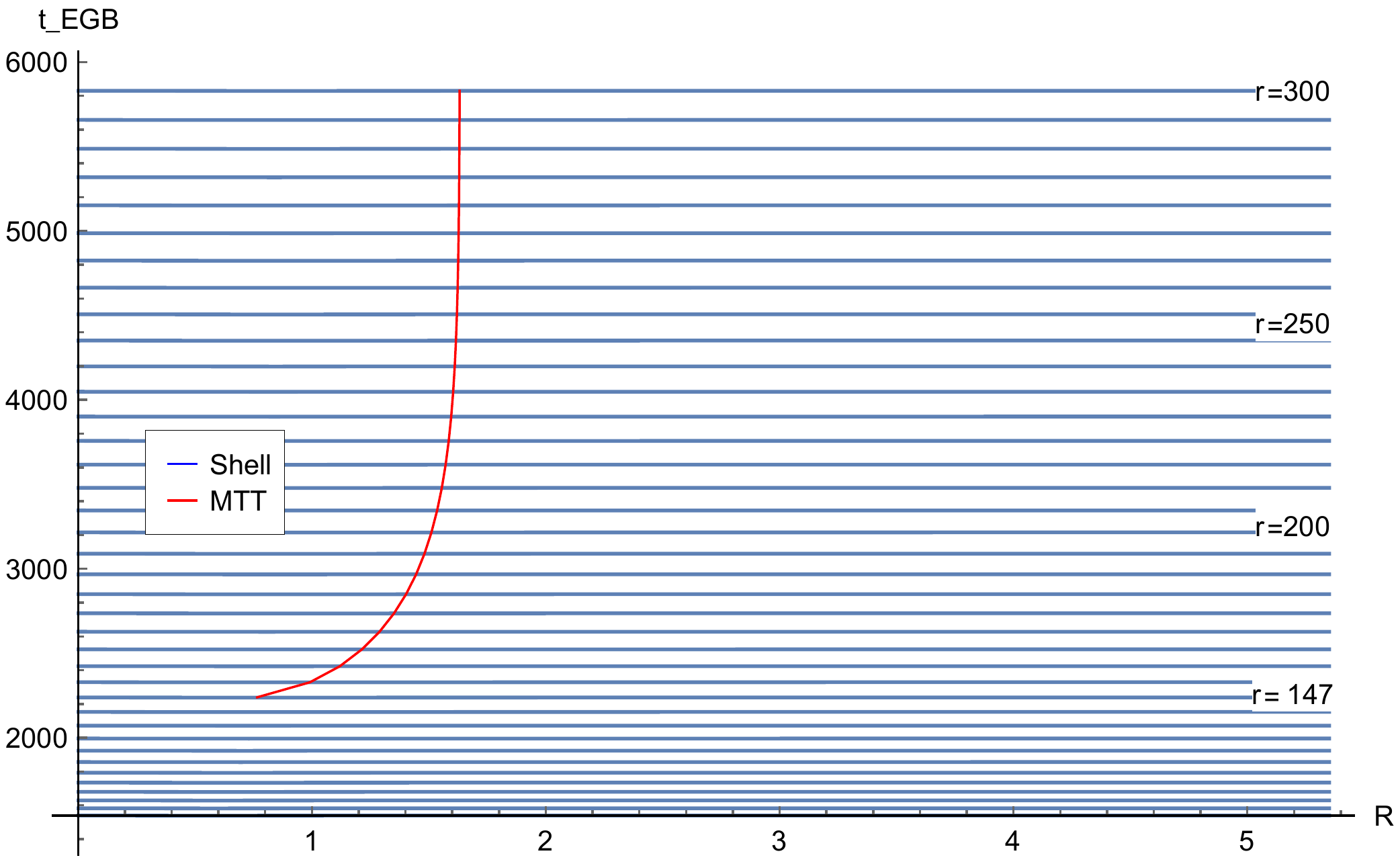}
		\caption{}
	\end{subfigure}
	\caption{The graphs show the (a) density distribution, (b) values of $C$,
		and (c) formation of MTT along with the shells  for $\lambda=0.6$. The MTT begins only after 
		sufficient mass has fallen in. The central central singularity is naked, and formation of trapped surfaces are delayed.
		The straight lines of MTT in (c), after the shell $r=250$, shows that the black hole
		has reached equilibrium state.}
	\label{fig:ltbkg0_example5}
\end{figure}
%
In our example,
we have chosen $r_{0}=100\, m_{0}$. Initially, the shells collapse
to the central singularity. There 
the singularity is naked and massive. However, only after the shell labelled by $r=140$ (approximately)
has collapsed, and the mass of the collapsed shells $\ge 0.3$,
 that the MTT begins from, remains spacelike until all the 
mass has fallen in. After the shell $r=300$, the MTT reaches equilibrium and becomes isolated and null,
see fig. \eqref{fig:ltbkg0_example5}(c). The signature of MTT may 
be verified from  figure \eqref{fig:ltbkg0_example5}b.

(ii) Two shells falling consecutively on a black hole:
Let us assume that a black hole of mass $M=1$ exists, upon which a density profile of the following form
falls:
\begin{equation}
\rho(r)=\frac{8\,(m_{0}/\,r_{0}^{3}\,)\,[(r/r_{0})-\sigma]^{\,2}}{[2\sigma +(3+2\sigma^{2})\sqrt{\pi}e^{\sigma^{2}}\{1+\erf(\sigma)\}]}
\, \exp[(2r/r_{0})\sigma-(r/r_{0})^{\,2}\,],
\end{equation}
where $m_{0}=M/2$ is the mass of the shell, $2r_{0}$ is the width of each shell.  
%
\begin{figure}[h!]
	\begin{subfigure}{.35\textwidth}
		\centering
		\includegraphics[width=\linewidth]{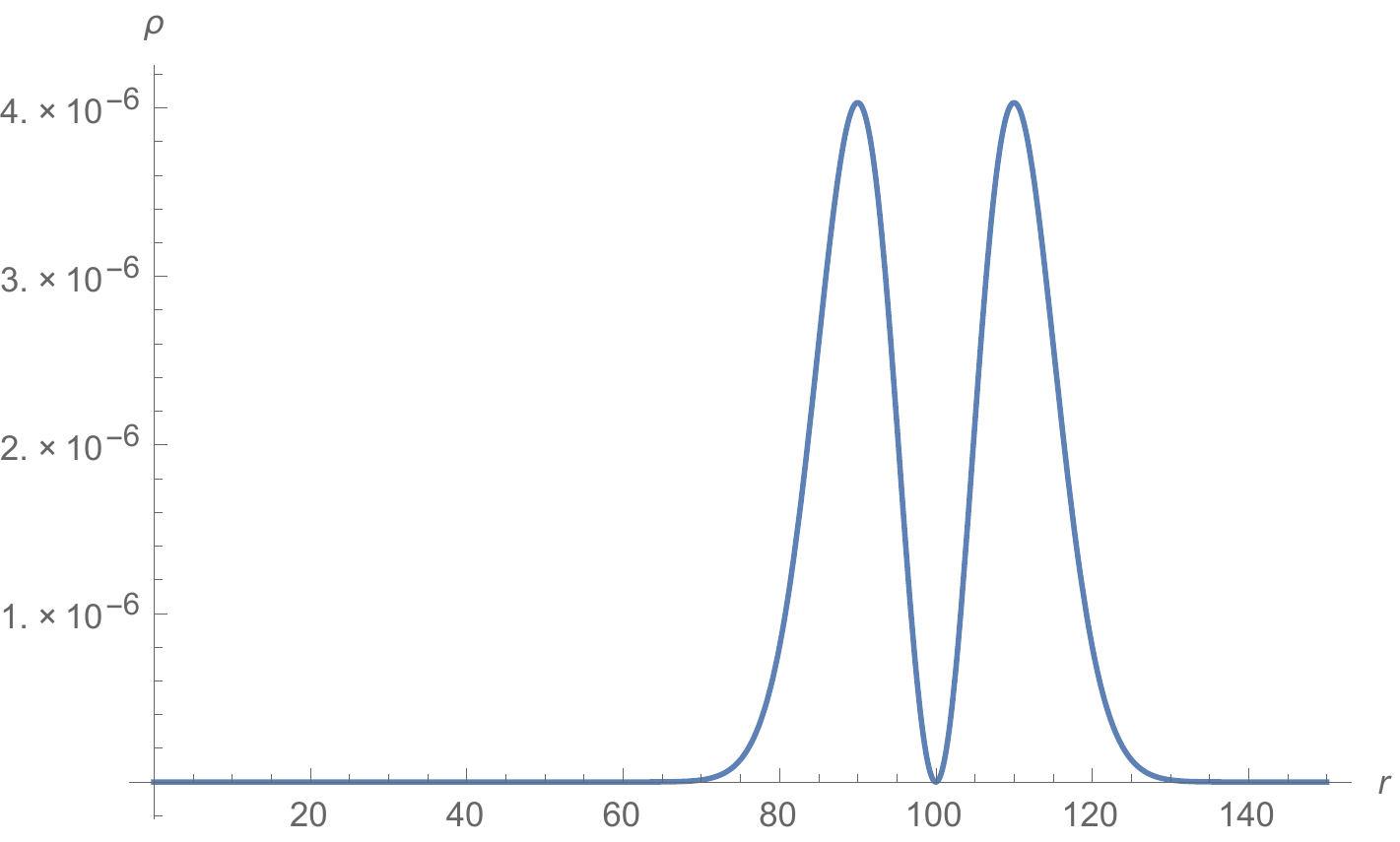}
		\caption{}
	\end{subfigure}
	\begin{subfigure}{.45\textwidth}
		\centering
		\includegraphics[width=\linewidth]{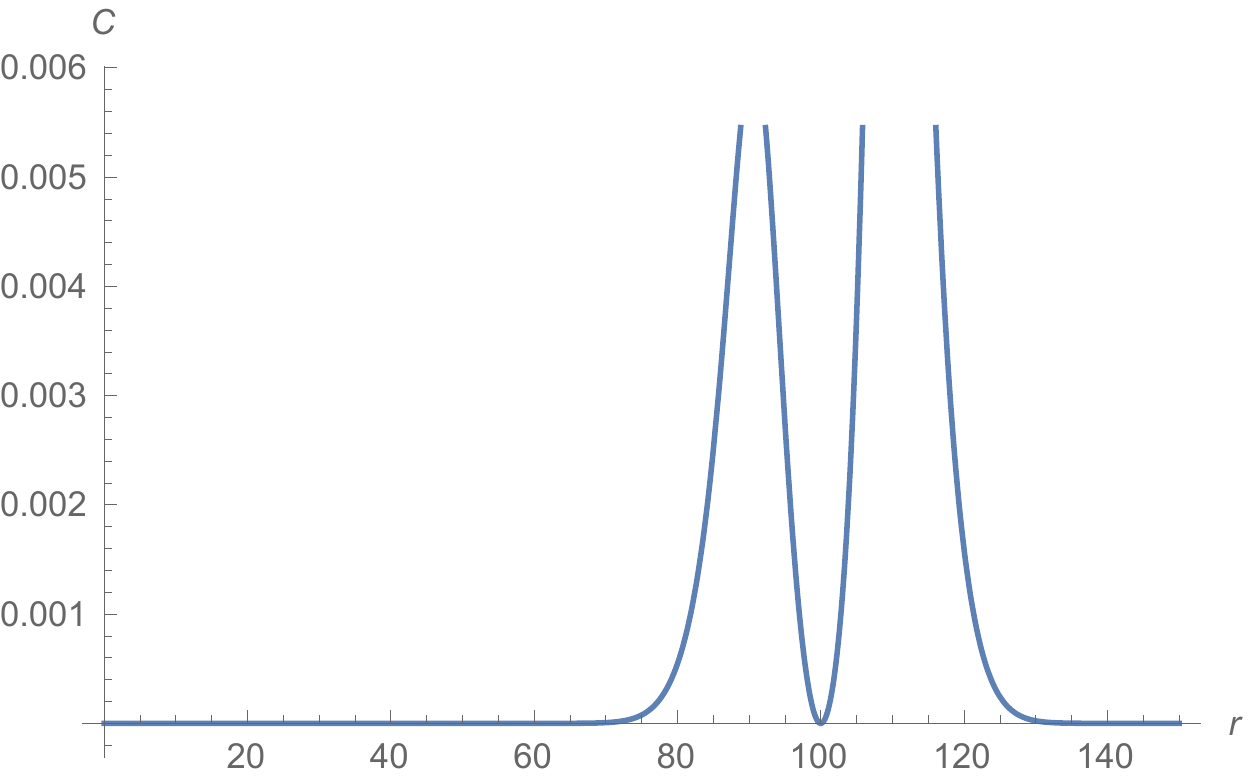}
		\caption{}
	\end{subfigure}
	\begin{subfigure}{.55\textwidth}
		\includegraphics[width=\linewidth]{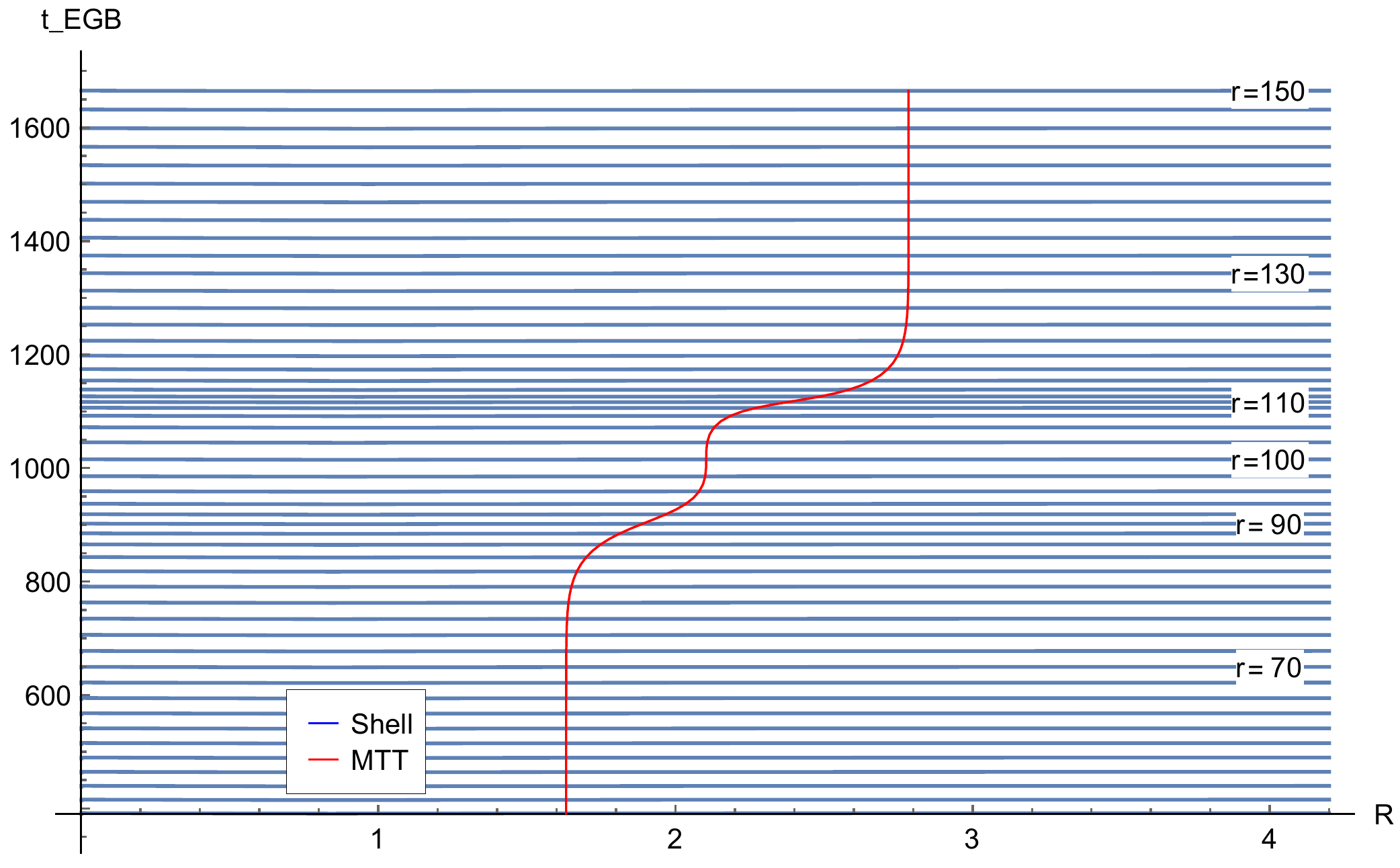}
		\caption{}
	\end{subfigure}
	\caption{The graphs show the (a) density distribution, (b) values of $C$,
		and (c) formation of MTT along with the shells which fall consecutively 
		on a black hole for $\lambda=0.6$. The MTT begins from $R=1.6$, where the previous EH is situated.
		The straight lines of MTT in (c) represents the isolated horizon phase.}
	\label{fig:ltbkg0_example6}
\end{figure}
%
Given the mass of initial black hole, the initial EH radius given by $1.6$. the mass
for each shell of radius $r (r> \bar{r})$ is then $m(r)=M+4\pi\int_{\bar{r}}^{r}\rho(\hat{r})
\hat{r}^{2}\, d\hat{r}$. The quantity $\sigma$ is a parameter which denotes the position where
the density vanishes. Here, we have used $M=1$, $r_{0}=10$ and $\sigma=10$. Not surprisingly,
as we have explained, there is no naked singularity here. Because of the presence of a black hole
on the initial Cauchy hypersurface, the singularity remains covered throughout. At around $r=70$,
the MTT starts to grow in a spacelike fashion and reaches approximately at $R=1.5$ at
approximately $t=1014$ when the $r=100$th shell falls. Note that at this time the $C$ vanishes
making the MTT null. This is expected since the density of the shell goes to zero here
(see figure \eqref{fig:ltbkg0_example6}a). Again, just as 
the next shell starts to fall, the MTT again begins to evolve in a spacelike fashion to reach
$R=2.8$ at $t=1664$ when the shell denoted by $r=150$ has fallen in.


\section{Censorship and visibility of singularities}
The eqn. \eqref{radius_mtt} is fundamental since it relates the radius of 
the collapsing matter sphere with the mass function, and specifically implies 
that a trapped region forms if :
\begin{eqnarray}
F(r)\ge R(r,t)+\frac{\lambda}{R(r,t)}~. 
\end{eqnarray}
The natural question is the following: For a large part of the trapped region, until $F(r)\ge 
2\sqrt{\lambda}$ the singularity is naked, and hence, this region violates the weak cosmic censorship.
Null rays emanating from this region reach the future null infinity, although they may be 
highly redshifted to be detected. The central singularity at $r=0$ on the other hand is a singularity
even after MTT has covered it. The question may raise is: does this singular point violate
the strong version of cosmic censorship? To answer this question, we shall apply the widely 
used root equation method \cite{Joshi:1987wg}. 

Let us begin by redefining $R(r,t)=r\,v(r,t)$, where
the function $v(r,t_{i})=1$, and $v(r,t_{s})=0$, where $t_{i/s}$ are the initial and the singularity
times respectively. The condition for collapse $\dot{R}(r,t) <0$, shall now be replaced by
the requirement $\dot{v}(r,t) <0$. 
From the equation \eqref{dotRk}, the equation for the function $v(r,t)$ is obtained to be:
\begin{equation}\label{vdoteqn}
\dot{v}(t,r)=-\left[-k(r)-\frac{1}{2\lambda}\left(v^{2}-\sqrt{v^{4}+4\lambda F(r)\, v}\right)\right]^{1/2}.
\end{equation}
The equation of the time curve of shells is obtained by integrating this equation with respect to $v$, leading to:
\begin{equation}\label{shell_curve_eqn}
t(v,r)=\int_{v}^{\,1} \frac{d\bar{v}}{\left[-k(r)-\frac{1}{2\lambda}\left(\bar{v}^{2}
-\sqrt{\bar{v}^{4}+4\lambda F(r)\, \bar{v}}\right)\right]^{1/2}}~~.
\end{equation}
The singularity time ($t_{s}$) is determined by using the value of $v=0$ in the above integral.   
\begin{equation}
t_{s}(r)=\int_{0}^{\,1} \frac{d\bar{v}}{\left[-k(r)-\frac{1}{2\lambda}\left(\bar{v}^{2}
-\sqrt{\bar{v}^{4}+4\lambda F(r)\, \bar{v}}\right)\right]^{1/2}}~~,
\end{equation}
whereas the time for the shells to reach the MTT, denoted by $t_{\scaleto{MTT}{4pt}}$ is 
obtained by using the appropriate time values
of the shell labelled by $r$ to reach that particular MTT radius  
$R_{\scaleto{MTT}{4pt}}=rv_{\scaleto{MTT}{4pt}}$. This equation is given by:
\begin{equation}
t_{\scaleto{MTT}{4pt}}\,(r)=\int_{v_{\scaleto{\,MTT}{3pt}}}^{\,1} \frac{d\bar{v}}{\left[-k(r)-\frac{1}{2\lambda}\left(\bar{v}^{2}
-\sqrt{\bar{v}^{4}+4\lambda F(r)\, \bar{v}}\right)\right]^{1/2}}.
\end{equation}
A slight rearrangement of this equation lead to the following expression in terms of $t_{s}(r)$:
\begin{equation}
t_{\scaleto{MTT}{4pt}}\,(r)=t_{s}(r)-\int^{v_{\scaleto{\,MTT}{3pt}}}_{\,0} \frac{d\bar{v}}{\left[-k(r)-\frac{1}{2\lambda}\left(\bar{v}^{2}
-\sqrt{\bar{v}^{4}+4\lambda F(r)\, \bar{v}}\right)\right]^{1/2}}.
\end{equation}
This shows in a quite straightforward way that $t_{s}(r)$ is usually higher than
$t_{\scaleto{MTT}{4pt}}$ provided that the result of integration is positive.
We have shown in quite detail that unless $F\ge 2\sqrt{\lambda}$, no MTT forms,
and in this bound, when MTT is formed,  $t_{\scaleto{MTT}{4pt}}(r)< t_{s}(r)$. 

Now, using these equations let us inquire the visibility of the singularities, near 
the matter center $r=0$. Near $r=0$, all the functions considered above
may be taken to be a smooth function of $r$ (note for example that all the density
and mass functions considered in the previous sections have a smooth beheviour near $r=0$), and 
hence are Taylor expandable:
\begin{eqnarray}\label{Fk_taylor}
F(r)=F_{0}+F_{2}\, r^{2} +F_{4}\, r^{4} +\cdots , \\
k(r)=k_{0}+k_{2}\, r^{2} +k_{4}\, r^{4} +\cdots ,
\end{eqnarray}
where, $F_{i}$\,s and $k_{i}$\,s are constants, and only 
the even powers survive since the system is symmetric about $r=0$. Similarly, the equation for 
the time curve $t(v,r)$, may be expanded:
\begin{equation}
t(v,r)=t(v,0)+(1/2)\,r^{2}\, \chi_{2}(v) + \cdots ,
\end{equation}
where the function $\chi_{2}(v)$ is obtained from the eqn. \eqref{shell_curve_eqn}, and gives:
\begin{equation}
\chi_{2}\,(v)=\int_{v}^{\,1} \frac{k_{2}- F_{2}\,\left(\bar{v}^{2}
+4\lambda\, F_{0}\, \bar{v}^{-1}\right)^{-1/2}}{\left[-k_{0}-\frac{1}{2\lambda}\left(\bar{v}^{2}
-\sqrt{\bar{v}^{4}+4\lambda\, F_{0}\, \bar{v}}\right)\right]^{3/2}} \,\, d\bar{v}.
\end{equation}
For understanding the visibility of the singularities, we require future- directed outgoing
radial null geodesics to emerge from the singularity at $r=r_{s}=0$.
The fixed point method requires that we define a quantity
\begin{equation}
x=\frac{R(r,t)}{r^{q}}=\frac{v(r,t)}{r^{q-1}},
\end{equation}
where $q\ge 0$ is a constant. We use the root equation method, and accordingly,
near the central singularity, 
\begin{eqnarray}
x_{0}&=&\lim_{r\rightarrow \, 0} \frac{R}{r^{q}}
=\lim_{r\rightarrow \,0}\, \frac{1}{q\,r^{q-1}}\times \frac{dR}{dr}\nonumber\\
&=& \lim_{r\rightarrow \, 0}\,\left( \frac{v+rv^{\prime}}{q\,r^{q-1}}\right)\left(1+\frac{r\dot{v}}{\sqrt{1-kr^{2}}}\right)
\end{eqnarray}
Using the equations \eqref{vdoteqn}, and the expansions in \eqref{Fk_taylor}, we get 
the desired root equation, which is to be evaluated at $v=x_{0}\, r^{(q-1)}$: 
\begin{equation}
x_{0}= \lim_{r\rightarrow \, 0}\,\frac{1}{q\,r^{q-1}}\left[1-r\left(\frac{v\,F_{0}}{\lambda}\right)^{1/4}\right]\times\left[v+\left(\frac{v\, F_{0}}{\lambda}\right)^{1/4}
r^{2}\, \chi_{2}(0)\right].
\end{equation}
The finite positive root of this equation is obtained with $q=(11/3) >1$, such that
\begin{equation}
x_{0}^{3/4}=(3/8)\,\left(\frac{F_{0}}{\lambda}\right)^{4}\, \chi_{2}(0)\, .
\end{equation}
This implies that, near 
the central singularity, the future directed radial null geodesic behaves as $R(r,t)=x_{0}\, r^{11/3}$.
The density again diverges at $r=0$, since from eqn. \eqref{1eq1}, the expression for 
density contains $R^{\,\prime}(r,t)=(v+rv^{\prime})$. The expression for $v^{\prime}$ is given by:
\begin{equation}
v^{\prime}=\left(\frac{x_{0}\,F_{0}}{\lambda }\right)^{1/4}\, r^{5/3}\, \chi_{2}(0).
\end{equation}
Hence $r\,v^{\prime}\rightarrow \, 0$ as $r\,\rightarrow \, 0$, leading to the divergence of the 
density as well as the curvature along this null geodesic. This establishes that the singularity
at $r=0$ is at least locally naked, and hence the strong version of the cosmic censorship is violated.

\section{Discussion}
In this paper, we studied gravitational collapse of inhomogeneous dust in $4$d EGB gravity.
The matter is assumed to be, initially, either bounded or marginally bounded. 
The collapse phenomenon 
presents some surprising results which are absent in $4$d general relativity. Using a wide variety
of matter profiles, we show that if there are no black holes on the Cauchy hypersurface,
(i) the central singularity is initially naked: the formation of trapped surfaces are delayed until
mass contained inside the collapsed sphere exceeds $\sqrt{\lambda}$, where $\lambda$
is the GB coupling constant,
(ii) MTT begins non- centrally when the equality is satisfied, and (iii) the MTT covers
the singularity if $F(r)> 2\sqrt{\lambda}$. For the situation when there exists a black hole
on the initial data set, the central singularity is censored, and the MTT begins from 
the first collapsing shell itself. Thus this theory provides
a fertile ground to study effects of naked singularities.

Given these results, several queries arise. 
The first concerns the nature of naked singularity formed initially. 
For a large class of smooth matter profiles, one can obtain null geodesics emanating 
from the central singularity. For our case, initially for $F(r)< 2\sqrt{\lambda}$, 
the metric to which the naked singularity needs to be matched, is globally naked. Hence, until 
the mass function satisfies this
bound, the singularity is globally naked, and violates the weak censorship. On the other hand, if 
the singularity is locally naked, it may be considered to violate 
the strong version cosmic censorship. As we saw in the 
previous section, the central singularity is always locally naked, whether it is covered or not.
Indeed, when $F(r)\ge 2\sqrt{\lambda}$\, , the singularity violates the strong censorship.
However, we must also note that, the central singularity at the least, is not strong, which, according to 
the analysis carried out in \cite{tce, joshi}, is essential for proving non- extendibility of spacetimes.
The sufficient condition for a strong curvature singularity is 
that for at least one causal 
geodesic $\tau^{\mu}$, with affine parameter $s$, in the limit
to singularity must obey:
\begin{equation}
\lim_{s\rightarrow \, s_{0}}\, (s-s_{0})^{2}\, R_{\mu\nu}\, \tau^{\mu}\, \tau^{\nu} \, >\, 0,
\end{equation}
For a normalised radial timelike geodesic corresponding to the metric in eqn. \eqref{extmetric}, 
near the center, and with $\tau^{\mu}=\delta^{\mu}_{t}$, the above expression is zero. Hence,
in $4$d, the EGB theory, the singularity does not satisfy the strong curvature condition. 

The second is about the mass of the singularity.
Unlike GR, massive singularities are no unheard of in higher curvature theories. 
Our solutions give explicit constructions of these objects. How does one quantify them,
and what role shall they play in quantum theory? One also should determine 
the stability and experimental
signatures of such spacetime structures \cite{Aragon:2020qdc, Chowdhury:2022zqg}. These studies, with a detail description of the dynamics of the scalar field in the $4$d EGB action shall be carried out in 
the future.

\section*{Acknowledgements}
The author AC is supported through the DAE-BRNS project $58/14/25/2019$-BRNS,
and by the DST-MATRICS scheme of government of India through their grant MTR-$/2019/000916$.

%

\section*{Appendix 1: The field equations}
The Lagrangian in eqn. \eqref{trace_anomaly}, belongs to the class of most general 
gravity Lagrangians, linear in the curvature scalar $R$, quadratic in $\phi$, and containing terms with quartic derivatives \cite{Amendola:1993uh,Capozziello:1999uwa,Daniel:2007kk,Sushkov:2009hk,Granda:2009fh,Gao:2010vr,Granda:2011eh}. One further interesting point is that the kinetic term of 
the scalar is coupled
to the Einstein tensor. Such theories have been considered in great detail as useful models for both the
dark energy and the dark matter \cite{Granda:2009fh,Gao:2010vr,Granda:2011eh}. In the following we write the field equations for the Lagrangian in eqn.\eqref{trace_anomaly} \cite{Granda:2011eh,Ma:2020ufk}:
\begin{eqnarray*}\label{app_trace_anomaly}
L_{4}=\sqrt{-g} [R+\lambda \left\{ \phi\, \mathcal{G}+4G^{\mu\nu} \partial_{\mu}\phi\, \partial_{\nu}\phi
-4(\partial_{\mu}\phi)^{2}\, \Box \phi +2\,(\partial_{\mu}\phi)^{4}\right\}].
\end{eqnarray*}
The metric variation of the action gives the following terms. Note that no matter Lagrangian has been added here, however, if they are added, they would contribute an additional factor 
of $8\pi T^{\text{\,matter}}_{\mu\nu}$:
\begin{equation}\label{app_Einstein}
G_{\mu\nu}=\lambda T^{(1)}_{\mu\nu}+2\lambda T^{(2)}_{\mu\nu}-\lambda T^{(3)}_{\mu\nu},
\end{equation}
where $T^{(1)}_{\mu\nu}$ is due to the variation of $\sqrt{-g} (\phi\, \mathcal{G}\,)$, $T^{(2)}_{\mu\nu}$ is due to the variation of the quantity $\sqrt{-g} (G^{\mu\nu} \partial_{\mu}\phi\, \partial_{\nu}\phi)$, whereas
$T^{(3)}_{\mu\nu}$ is due to the variation of $\sqrt{-g} \left[-4(\partial_{\mu}\phi)^{2}\, \Box \phi +2\,(\partial_{\mu}\phi)^{4}\right]$. The quantities are listed below:
\begin{eqnarray}
T^{(1)}_{\mu\nu}&=&2\left[R\,\nabla_{\mu}\nabla_{\nu}\phi +2G_{\mu\nu}\Box \phi+2(R_{\mu\nu\lambda\sigma}-g_{\mu\nu}R_{\lambda\sigma})\,\nabla^{\lambda\sigma}\phi-4R_{\lambda(\mu}\nabla^{\lambda}\nabla_{\nu)\phi}\right] \nonumber\\
&& ~~ - \phi\, H_{\mu\nu},
\end{eqnarray}
where, $H_{\mu\nu}$ is given by eqn. \eqref{hab_eqn}, and
\begin{eqnarray}
&&T^{(2)}_{\mu\nu}=G_{\mu\nu}(\nabla_{\rho}\phi)^{2} +R\nabla_{\mu}\phi\nabla_{\nu}\phi-4R_{(\mu}{}^{\rho}\nabla_{\nu)}\phi\nabla_{\rho}\phi+2(\nabla_{\mu}\nabla_{\nu}\phi)\Box\phi-2(\nabla_{\rho}\nabla_{\mu}\phi)(\nabla^{\rho}\nabla_{\nu}\phi)\nonumber\\
&&~~~~~~ -g_{\mu\nu}\,(\Box\phi)^{2}+2g_{\mu\nu}R^{\rho\sigma}\nabla_{\rho}\phi\nabla_{\sigma}\phi +g_{\mu\nu}
(\nabla_{\rho}\nabla_{\sigma}\phi)(\nabla^{\rho}\nabla^{\sigma}\phi)+2R_{\mu\nu\lambda\sigma}\nabla^{\lambda}\phi\nabla^{\sigma}\phi,
\end{eqnarray}
\begin{eqnarray}
&&T^{(3)}_{\mu\nu}=4\nabla_{\mu}\phi\nabla_{\nu}\phi\, (\nabla_{\rho}\phi)^{2}-g_{\mu\nu}\{(\nabla_{\rho}\phi)^{2}\}^{2}-4(\nabla_{\mu}\nabla_{\nu}\phi)\Box\phi-4g_{\mu\nu}(\nabla^{\rho}\nabla^{\sigma}\phi)\nabla_{\rho}\phi\nabla_{\sigma}\phi\nonumber\\
&&~~~~~~~~~~~~~~~~~~~~~~~ +8\nabla^{\rho}\nabla_{(\mu}\phi\, \nabla_{\nu)}\phi\nabla^{\rho}\phi.
\end{eqnarray}

The equation of motion of the scalar field is obtained by variation over $\phi$. This gives 
the following equation:
\begin{eqnarray}\label{app_scalar_eqn}
&&\Box\phi\,\left[(R/2)+\Box \phi +8\, (\nabla_{\mu}\phi)^{2}\right]
=R_{\mu\nu}\left[\nabla^{\mu}\phi\nabla^{\nu}\phi+\nabla^{\mu}\nabla^{\nu}\phi\right]+2(\nabla_{\mu}\nabla_{\nu}\phi)(\nabla^{\mu}\phi)(\nabla^{\nu}\phi)\nonumber\\
&&~~~~~~~~~~~~~~~~~~
~~+(\nabla_{\mu}\nabla_{\nu}\phi)^{2}+\mathcal{G}.
\end{eqnarray}
The trace of the Einstein equation eqn. \eqref{app_Einstein} leads to the following constraint:
\begin{equation}
R+ (\text{EoM of scalar field})+(\lambda/2)\mathcal{G}=0,
\end{equation}
and on-shell, this leads to the equation relating the Ricci scalar and the GB scalars the Lagrangian of 
the theory.
\section*{Appendix 2: Equations for shell collapse}
Here, we give the detailed expressions for the time of formation of the MTT for each collapsing shell, and the 
singularity formation time. These equations have been used in the section $3.2$, for plots of bounded collapse of matter configurations.
\begin{eqnarray}\label{appeq1}
t_{c}&=& \sqrt{2\,\lambda}\left[I_{1}+I_{2}+\frac{r^{\frac{3}{2}}\tan^{-1}\left[ \frac{r\,x}{2\,F\,\lambda}\right]^{1/2}}{2\,\sqrt{2\,F\,\lambda}}+ \frac{I_{3}\,I_{13}\,(-1)^\frac{1}{3}\,Q_{3} \, \mathrm{EllipticF} \left[\tan^{-1}[I_{6}],I_{7}\right]}{I_{15}\,I_{14}}\nonumber\right.\\&-& \left.  \frac{I_{3}}{I_{15}\,I_{12}}\left\{I_{4}+ I_{5}\,\frac{(-1)^{\frac{1}{3}}+(-1)^{\frac{2}{3}} }{1+(-1)^{\frac{1}{3}}} \,\mathrm{EllipticE} \left[\tan^{-1}[I_{6}],I_{7}\right]\nonumber
\right.\right.\\
&+&\left.\left.
\frac{I_{5}\,I_{8}\,\mathrm{EllipticF} \left[\tan^{-1}[I_{6}],I_{7}\right]}{I_{9}}  -\frac{I_{5}\,I_{10}\,\mathrm{EllipticPi} \left[I_{11},\tan^{-1}[I_{6}],I_{7}\right]}{\{2F\lambda\}/r-Q_{3}}\right\} \nonumber
\right.\\
&+&\left.\frac{I_{3}\,I_{13}\,\left\{1+(-1)^{\frac{1}{3}}\right\}\,Q_{3}\,\mathrm{EllipticPi} \left[I_{11},\tan^{-1}[I_{6}],I_{7}\right]}{I_{14}\,I_{15}} 
\right]-\left[\, \cdots \,\right]_{R=r}~~ ,
\end{eqnarray}
where $\left[\, \cdots \,\right]_{R=r}$ is the same expression as in the previous braces, however, now determined at $R=r$.
\begin{eqnarray}\label{appeq2}
t_{{\scaleto{MTT}{3pt}}}&=& \sqrt{2\,\lambda}\left[I_{1}+I_{2}+\frac{r^{\frac{3}{2}}\tan^{-1}\left[ \frac{r\,x}{2\,F\,\lambda}\right]^{1/2}}{2\,\sqrt{2\,F\,\lambda}}+ \frac{I_{3}\,I_{13}\,(-1)^\frac{1}{3}\,Q_{3} \, \mathrm{EllipticF} \left[\tan^{-1}[I_{6}],I_{7}\right]}{I_{15}\,I_{14}}\nonumber\right.\\&-& \left.  \frac{I_{3}}{I_{15}\,I_{12}}\left\{I_{4}+ I_{5}\,\frac{(-1)^{\frac{1}{3}}+(-1)^{\frac{2}{3}} }{1+(-1)^{\frac{1}{3}}} \,\mathrm{EllipticE} \left[\tan^{-1}[I_{6}],I_{7}\right]\nonumber
\right.\right.\\
&+&\left.\left.
\frac{I_{5}\,I_{8}\,\mathrm{EllipticF} \left[\tan^{-1}[I_{6}],I_{7}\right]}{I_{9}}  -\frac{I_{5}\,I_{10}\,\mathrm{EllipticPi} \left[I_{11},\tan^{-1}[I_{6}],I_{7}\right]}{\{2F\lambda\}/r-Q_{3}}\right\} \nonumber
\right.\\
&+&\left.\frac{I_{3}\,I_{13}\,\left\{1+(-1)^{\frac{1}{3}}\right\}\,Q_{3}\,\mathrm{EllipticPi} \left[I_{11},\tan^{-1}[I_{6}],I_{7}\right]}{I_{14}\,I_{15}} 
\right]_{R_{\scaleto{MTT}{3pt}}}-\left[\, \cdots \,\right]_{R=r}	~~ .  
\end{eqnarray}
The functions 
defined above, as $\mathrm{Elliptic \cdots}$ are the elliptic functions of various kinds \cite{}.
The variables of these functions are expressions given by:
\begin{eqnarray*}
&&Q_{1}=\left(x+\frac{2\,F\,\lambda}{r} \right), ~~Q_{2}=(4F\lambda)^{1/3}, ~~
Q_{3}=(2F^{2}\lambda^{2})^{1/3}\\
&&Q_{4}=\sqrt{Q_{2}-r}\hspace{0.2cm}; \hspace{0.5cm} Q_{5}=\sqrt{Q_{2}+{(-1)}^{\frac{1}{3}} r}.
\end{eqnarray*}
The expressions $I_{i}$ in the equations \eqref{appeq1}, and \eqref{appeq2}, are given by:
\begin{eqnarray*}
I_{1}&=&\frac{r\,\sqrt{x}}{2\,Q_{1}}\hspace{0.2cm}; \hspace{0.5cm} I_{2}=-\frac{r\,\sqrt{x}\sqrt{2F^2\lambda^2-Q_{1}^3}}{2\sqrt{2}F\lambda\,Q_{1}}\hspace{0.2cm}; \hspace{0.5cm}
I_{3}=\frac{r\,x\,\sqrt{Q_{1}^3-2F^2\lambda^2}}{2\,Q_{1}\,\left(4F^2\lambda^2+Q_{1}^{3} \right)^{-1}}\\
I_{4}&=&\frac{x\,\left(Q_{1}+{(-1)}^{\frac{1}{3}}\,Q_{3} \right)}{\left(Q_{1}-{(-1)}^{\frac{2}{3}}\,Q_{3} \right)^{-1}}\hspace{0.1cm}; \hspace{0.1cm}\hspace{0.1cm}
I_{5}= \frac{{(-1)}^{\frac{3}{4}}\,(1+{(-1)}^{\frac{1}{3}})}{\sqrt{2}\,{3}^{\frac{1}{4}}\,Q_{5}\,Q_{2}^{-1}\,Q_{4}^{-1}}\,\frac{\sqrt{Q_{1}-Q_{3}}\, \sqrt{Q_{1}-{(-1)}^{\frac{2}{3}}\,Q_{3}}}{\left[\frac{2\,F\,\lambda}{r}+{(-1)}^{\frac{1}{3}}\,Q_{3}\right]^{\frac{-1}{2}}}\\
I_{6}&=& \frac{Q_{4}\,\sqrt{Q_{1}+{(-1)}^{\frac{1}{3}}\,Q_{3}}}{Q_{5}\,\sqrt{Q_{1}-Q_{3}}}\hspace{0.1cm}; \hspace{0.1cm}
I_{7}=-\frac{{(-1)}^{\frac{2}{3}}\,(-1+{(-1)}^{\frac{2}{3}})\,Q_{5}^2}{Q_{4}^2}
\end{eqnarray*}
\begin{eqnarray*}
I_{8}&=&Q_{3}^{2} \left[(-1+{(-1)}^{\frac{1}{3}})-\frac{Q_{2}}{r}\,(1+{(-1)}^{\frac{1}{3}}) \right]\hspace{0.1cm}; \hspace{0.1cm}
I_{9}= -(1+{(-1)}^{\frac{1}{3}})\, Q_{3}^{2} \left(1-\frac{Q_{2}}{r} \right)\\
I_{10}&=&-\frac{2\,F\,\lambda}{r}\hspace{0.1cm}; \hspace{0.5cm}
I_{11}= \frac{Q_{5}^{2}}{Q_{4}^{2}}\hspace{0.1cm}; \hspace{0.5cm}
I_{12}= \sqrt{x}\,\sqrt{Q_{1}^{3}-2\,F^2\,\lambda^2}\\
I_{13}&=& \frac{{(-1)}^{\frac{5}{12}}\,\sqrt{2x}\,(1+{(-1)}^{\frac{1}{3}})}{{3}^{\frac{1}{4}}\,Q_{5}\,F^{-1}\lambda^{-1}\,Q_{4}^{-1}}\,\frac{\sqrt{Q_{1}-Q_{3}}\, \sqrt{Q_{1}-{(-1)}^{\frac{2}{3}}\,Q_{3}}}{\left[Q_{1}+{(-1)}^{\frac{1}{3}}\,Q_{3}\right]^{\frac{-1}{2}}\,\left[\frac{2\,F\,\lambda}{r}+{(-1)}^{\frac{1}{3}}\,Q_{3}\right]^{\frac{-1}{2}}}\\
I_{14}&=&\sqrt{x} Q_{3}^{2}\,(1+{(-1)}^{\frac{1}{3}})\left(1-\frac{Q_{2}}{r}\right) \sqrt{Q_{1}^{3}-2\,F^2\,\lambda^2}\\
I_{15}&=& \sqrt{2}\,\frac{F\,\lambda\,r\,x}{Q_{1}} \sqrt{2\,F^2\,\lambda^2-Q_{1}^{3}} \left(4\,F^2\lambda^2+Q_{1}^3\right)
\end{eqnarray*}
These equations have been heavily used to plot the MTTs for various matter field models.

\section{Appendix 3: Matching conditions}
The interior LTB metric of the spacetime $\mathcal{M^{-}}$ is given by eqn. \eqref{intmetric}.
The metric
of the external spacetime $\mathcal{M_{+}}$ is eqn.\eqref{extmetric}.
The matching is to be carried out at the timelike hypersurface $\Sigma$ given by $r_{b}$.
Let us denote the coordinates on this surface $\Sigma$ to be $(\tau, \theta, \phi, \psi)$.  
From $\mathcal{M^{-}}$, we can write down the surface $\Sigma$ as
$f_{-}(r,t)=r-r_{b}=0$, and hence, the induced metric on $\Sigma$ is
\begin{eqnarray}
ds^{2}_{-}&=&-d\tau^{2}+ r_{b}^{2}\, d \Omega_{2}\,.
\label{M_1}
\end{eqnarray}
From the point of view of the exterior spacetime, the hypersurface
may be described by $r=\bar{R}_{\Sigma}(\tau)$ and $t=T_{\Sigma}(\tau)$, with no change in
the angular variables.
The line element of the hypersurface is then given by
\begin{eqnarray}
ds^{2}_{+}&=&-\left[f(\bar{R}_{\Sigma})\dot{T}_{\Sigma}^2-f(\bar{R}_{\Sigma})^{-1}\dot{\bar{R}}_{\Sigma}^2\right]d\tau^2
+ \bar{R}_{\Sigma}(\tau)^2 d \Omega_{2}\,, \label{m11+}
\end{eqnarray}
where the dots imply derivative with respect to $\tau$.

The induced metric in equations in $(\ref{M_1})$ and $(\ref{m11+})$ must have matched metric functions.
This implies that:
\begin{eqnarray}\label{metric_matching}
f(\bar{R}_{\Sigma})\dot{T}_{\Sigma}^2-f(\bar{R}_{\Sigma})^{-1}\dot{\bar{R}}_{\Sigma}^2=1
\end{eqnarray}

Now, let $u^{\mu}$ and $n^{\mu}$ denote the velocity of the matter variables and the
normal to the $\Sigma$ respectively. They must satisfy the conditions
$u^{\mu}u_{\mu}=-1$, $n^{\mu}n_{\mu}=1$, whereas, $u^{\mu}n_{\mu}=0$. From the
interior spacetime, the expressions of these vectors is easily obtained:
\begin{equation}
u^{\mu}=\delta^{\mu}_{0}\equiv (\partial_{\tau})^{\mu},
\, \qquad  \, n_{\mu}=\frac{R^{\, \prime }}{\sqrt{1-k(r)}}\,(dr)_{\mu} .
\end{equation}
From the exterior spacetime, these vectors are also obtained similarly to give:
\begin{equation}
u^{\mu}=\dot{T}_{\Sigma}\,\,(\partial_{\tau})^{\mu} + \dot{\bar{R}}_{\Sigma}\,(\partial_{r})^{\mu} ,
\, \qquad  \, n_{\mu}=-\dot{\bar{R}}_{\Sigma}\,(d\tau)_{\mu}
+ \dot{T}_{\Sigma}\,\,(d{r})_{\mu} .
\end{equation}

The extrinsic curvatures are easily determined from these normals for the exterior as well the interior
spacetimes:
\begin{eqnarray}
&K^{-}_{\tau\tau}=0, \,\, & \qquad K^{-}_{\theta\theta}= \bar{R}_{\Sigma}\,\sqrt{1-k(r_{b})}\\
 &K^{+}_{\tau\tau}=\dot{\bar{R}}_{\Sigma}^{-1}[\dot{f}(\bar{R}_{\Sigma})\, \dot{T}_{\Sigma}
+{f}(\bar{R}_{\Sigma})\, \ddot{T}_{\Sigma}\, ] \,
 &\qquad K^{+}_{\theta\theta}= \bar{R}_{\Sigma}\,{f}(\bar{R}_{\Sigma})\, \dot{T}_{\Sigma}.
\end{eqnarray}
The $K_{\theta\theta}$ equations imply the following relation:
\begin{equation}
\frac{dT_{\Sigma}}{d\tau}=\frac{\sqrt{1-k(r_{b})}}{f(\bar{R}_{\Sigma})},
\end{equation}
whereas the equation (\ref{metric_matching}) gives the following equation for the function
$\bar{R}_{\Sigma}$ :
\begin{equation}
\frac{d\bar{R}_{\Sigma}}{d\tau}=[1-k(r_{b})-f(\bar{R}_{\Sigma})]^{1/2}.
\end{equation}
This implies that the following relation hold good:
\begin{equation}
\left(\frac{d\bar{R}_{\Sigma}}{d\tau}\right)^{2}=-k(r_{b})
+\frac{\bar{R}_{\Sigma}^{2}}{4\lambda}\left[1 \mp \sqrt{1+\frac{8\lambda M}{\bar{R}_{\Sigma}^{3}}}\right].
\end{equation}
A simple comparison with equation \eqref{dotRk}) implies that the condition $2M=F(r_{b})$
must be satisfied at the boundary.


\end{document}